\newif\ifarxiv
\arxivfalse
\newif\ifjaiarxiv
\jaiarxivtrue

\ifarxiv
  \documentclass[times]{aastex61}
  \newcommand\fullfiguresize{0.8}
  \newcommand\mediumfiguresize{0.5}
  \newcommand\smallfiguresize{0.5}
\else
  \documentclass[a4paper]{ws-jai}
  \ifjaiarxiv
    \newcommand\fullfiguresize{0.8}
    \newcommand\mediumfiguresize{0.5}
    \newcommand\smallfiguresize{0.5}
  \else
    \newcommand\fullfiguresize{1.0}
    \newcommand\mediumfiguresize{0.8}
    \newcommand\smallfiguresize{0.6}
  \fi
  \normalsize
\fi
\usepackage[utf8]{inputenc}
\newif\ifclean
\cleantrue
\ifarxiv
  \cleantrue
\fi
\ifjaiarxiv
  \cleantrue
\fi

\usepackage{calc}
\ifclean%Define clean to use todo's
  \ifarxiv
  
  \else
    \ifjaiarxiv
      \newlength\hlen
      \hoffset=\hlen
      \newlength\len
      \oddsidemargin=\len
      \evensidemargin=\len
    \else
      \newlength\len
    \fi
  \fi
\else
  \newlength\len
  \oddsidemargin=\len
  \evensidemargin=\len
  \marginparwidth=\len
\fi
\usepackage{booktabs}
\usepackage[flushleft]{threeparttable}
\usepackage{comment}
\usepackage{courier}

\usepackage[breaklinks,linktoc=all,colorlinks,urlcolor=blue,citecolor=blue,linkcolor=blue]{hyperref}

\usepackage{enumitem}

\ifarxiv

\else
  \setlist[itemize]{leftmargin=5.5mm}
\fi

\usepackage{array}

\usepackage[frozencache=true,cachedir=.]{minted}

\usepackage{wrapfig}

\ifarxiv
  \usepackage{savesym}
  \savesymbol{tablenum}
  \usepackage{siunitx}
  \restoresymbol{SIX}{tablenum}
\else
  \usepackage{siunitx}
\fi

\ifclean%Define clean to use todo's
  \usepackage[colorinlistoftodos,disable]{todonotes}
\else
  \usepackage[colorinlistoftodos]{todonotes}
\fi
\newcommand{\todofigure}{\todo[color=yellow!40]}

\newcommand{\tododetail}{\todo[color=green!40]}

\usepackage{relsize}
\newcommand\cpp{C\nolinebreak[4]\hspace{-.05em}\raisebox{.4ex}{\relsize{-3}{\textbf{++}}} }
\newcommand\cppnospace{C\nolinebreak[4]\hspace{-.05em}\raisebox{.4ex}{\relsize{-3}{\textbf{++}}}}
\usepackage{arydshln}

\usepackage{tikz}
\usepackage{listings}
\usetikzlibrary{decorations.pathreplacing,calc}
\newcommand{\tikzmark}[1]{\tikz[overlay,remember picture] \node (#1) {};}
\newcommand*{\AddNote}[4]{%
    \begin{tikzpicture}[overlay, remember picture]
        \draw [decoration={brace,amplitude=0.5em},decorate,ultra thick,blue]
            ($(#3)!([yshift=1.5ex]#1)!($(#3)-(0,1)$)$) --  
            ($(#3)!(#2)!($(#3)-(0,1)$)$)
                node [align=center, text width=2.5cm, pos=0.5, anchor=west] {#4};
    \end{tikzpicture}
}%

\usepackage{inconsolata}

\usepackage{ragged2e} %Causes text to break inside margins, e.g., for labels

\usepackage{cleveref} %Must be loaded as late as possible

\ifclean

\else
  \usepackage[notref,notcite]{showkeys} %Displays figure labels. Turn off (with clean) for final draft

  \renewcommand*{\showkeyslabelformat}[1]{%Causes text to break inside margins, e.g., for labels
    \expandafter\def\expandafter\UrlBreaks\expandafter{\UrlBreaks%  save the current one
    \do\a\do\b\do\c\do\d\do\e\do\f\do\g\do\h\do\i\do\j%
    \do\k\do\l\do\m\do\n\do\o\do\p\do\q\do\r\do\s\do\t%
    \do\u\do\v\do\w\do\x\do\y\do\z\do\A\do\B\do\C\do\D%
    \do\E\do\F\do\G\do\H\do\I\do\J\do\K\do\L\do\M\do\N%
    \do\O\do\P\do\Q\do\R\do\S\do\T\do\U\do\V\do\W\do\X%
    \do\Y\do\Z}%
    \fbox{\vbox{\hsize=1cm\RaggedRight\noindent\normalfont\small\url{#1}\par}}}

  \makeatletter
    \SK@def\Cref#1{\SK@\SK@@ref{#1}\SK@Cref{#1}}%Displays figure labels for cref
  \makeatother
\fi

\definecolor{codegray}{gray}{0.885}
\newmintinline[python]{python}{bgcolor=codegray,fontsize=\small}

\ifarxiv
  \shorttitle{Bifrost: High-Throughput Stream Processing in Astronomy}
  \shortauthors{Cranmer et al.}
\fi

\ifarxiv
  
\begin{document}
\fi

\title{Bifrost: a Python/\cpp Framework\\
for High-Throughput
Stream Processing in Astronomy
}

\ifarxiv
  \author{Miles D. Cranmer}
  \affiliation{Harvard-Smithsonian Center for Astrophysics, 60 Garden Street, Cambridge, MA 02138, USA}
  \affiliation{Department of Physics \& McGill Space Institute, McGill University, Montreal, QC H3W 2C4, Canada}

  \author{Benjamin R. Barsdell}
  \affiliation{Harvard-Smithsonian Center for Astrophysics, 60 Garden Street, Cambridge, MA 02138, USA}
  \affiliation{NVIDIA Corporation, 2701 San Tomas Expressway, Santa Clara, CA 95050, USA}

  \author{Danny C. Price}
  \affiliation{Harvard-Smithsonian Center for Astrophysics, 60 Garden Street, Cambridge, MA 02138, USA}
  \affiliation{University of California, Berkeley, Campbell Hall 339, Berkeley CA 94720, USA}

  \author{Jayce Dowell}
  \affiliation{Department of Physics and Astronomy, University of New Mexico, Albuquerque, NM 87131, USA}

  \author{Hugh Garsden}
  \affiliation{Harvard-Smithsonian Center for Astrophysics, 60 Garden Street, Cambridge, MA 02138, USA}

  \author{Veronica Dike}
  \affiliation{Department of Physics and Astronomy, University of New Mexico, Albuquerque, NM 87131, USA}

  \author{Tarraneh Eftekhari}
  \affiliation{Harvard-Smithsonian Center for Astrophysics, 60 Garden Street, Cambridge, MA 02138, USA}

  \author{Alexander M. Hegedus}
  \affiliation{Department of Atmospheric, Oceanic and Space Sciences, University of Michigan, Ann Arbor, MI 48109, USA}

  \author{Joseph Malins}
  \affiliation{Department of Physics and Astronomy, University of New Mexico, Albuquerque, NM 87131, USA}

  \author{Kenneth S. Obenberger}
  \affiliation{Air Force Research Laboratory Space Vehicles Directorate, 3550 Aberdeen Ave. SE, Kirtland AFB, NM 87117, USA}

  \author{Frank Schinzel}
  \affiliation{National Radio Astronomy Observatory, Socorro, NM 87801, USA}
  \affiliation{Department of Physics and Astronomy, University of New Mexico, Albuquerque, NM 87131, USA}

  \author{Kevin Stovall}
  \affiliation{National Radio Astronomy Observatory, Socorro, NM 87801, USA}
  \affiliation{Department of Physics and Astronomy, University of New Mexico, Albuquerque, NM 87131, USA}

  \author{Gregory B. Taylor}
  \affiliation{Department of Physics and Astronomy, University of New Mexico, Albuquerque, NM 87131, USA}

  \author{Lincoln J. Greenhill}
  \affiliation{Harvard-Smithsonian Center for Astrophysics, 60 Garden Street, Cambridge, MA 02138, USA}

\else
  \author{
  Miles D. Cranmer$^{\ast, \dagger, \#}$,
  Benjamin R. Barsdell$^{\ast, \ddagger}$,
  Danny C. Price$^{\ast, \S}$,
  Jayce Dowell$^{\P}$,
  Hugh Garsden$^{\ast}$,\\
  %Start alphabetical:
  Veronica Dike$^{\P}$,
  Tarraneh Eftekhari$^{\ast}$,
  Alexander M. Hegedus$^{\parallel}$,
  Joseph Malins$^{\P}$,
  Kenneth S. Obenberger$^{\ast \ast}$,\\
  Frank Schinzel$^{\dagger \dagger, \P}$,
  Kevin Stovall$^{\dagger \dagger, \P}$,
  Gregory B. Taylor$^{\P}$,
  %End alphabetical:
  and Lincoln J. Greenhill$^{\ast}$\\ \hfill \\
  \textit{Draft version August 1, 2017.
  Submitted to the Journal of Astronomical Instrumentation.}
  }

  \address{
  %\flushleft
  %\centering
  \begin{centering}
  \begin{tabular}{@{}l@{}}
  \\
    $^\ast$\ \  Harvard-Smithsonian Center for Astrophysics,\\
    \ \ \ \ 60 Garden Street,
    Cambridge, MA 02138, USA\\
    $^\dagger$\ \  Department of Physics \& McGill
    Space Institute, McGill University,\\
    \ \ \ \ Montreal, QC H3W 2C4, Canada\\
    $^\ddagger$\ \  NVIDIA Corporation, 2701 San Tomas Expressway, Santa Clara, CA 95050, USA\\
    $^\S$\ \  University of California, Berkeley,
    Campbell Hall 339, Berkeley CA 94720, USA\\
    $^\P$\ \  Department of Physics and Astronomy, University of New Mexico,\\
    \ \ \ \ Albuquerque, NM 87131, USA\\
    $^\parallel$\ \ Department of Atmospheric, Oceanic and Space Sciences,
    University of Michigan,\\
    \ \ \ \ Ann Arbor, MI 48109, USA\\
    $^{\ast \ast}$ Air Force Research Laboratory Space Vehicles Directorate,\\
    \ \ \ \ 3550 Aberdeen Ave. SE, Kirtland AFB, NM 87117, USA\\
    $^{\dagger \dagger}$ National Radio Astronomy Observatory, Socorro, NM 87801, USA\\
  \end{tabular}
  \end{centering}
  }
\fi

\ifarxiv

\else %Goes after if not arxiv
  \begin{document}
\fi

%\catchline{}{}{}{}{} % Publisher's Area please ignore

\ifarxiv

\else
  \maketitle
\fi

\ifjaiarxiv
  \newcommand\blfootnote[1]{%
    \begingroup
    \renewcommand\thefootnote{}\footnote{#1}%
    \addtocounter{footnote}{-1}%
    \endgroup
  }
  \blfootnote{$^\#$Corresponding author: \href{mailto:miles.cranmer@cfa.harvard.edu}{miles.cranmer@cfa.harvard.edu}}
\fi

%For now ambiguity with affiliation numberings
%\renewcommand*{\thefootnote}{\fnsymbol{footnote}}
\renewcommand*{\thefootnote}{\arabic{footnote}}

\begin{abstract}

Radio astronomy observatories with high throughput back end instruments require real-time
data processing. While computing hardware continues to advance rapidly,
development of real-time processing pipelines remains difficult and time-consuming,
which can limit scientific productivity.  Motivated by
this, we have developed Bifrost: an open-source software framework for
rapid pipeline development\footnote{\url{https://github.com/ledatelescope/bifrost}}. 
Bifrost combines a high-level Python interface with highly efficient
reconfigurable data transport and a library of computing blocks
for CPU and GPU processing. The framework is
generalizable, but initially it emphasizes the needs of high-throughput
radio astronomy pipelines, such as the ability to process data
buffers as if they were continuous streams, the capacity
to partition processing into distinct data sequences (e.g., separate
observations), and the ability to extract specific intervals from
buffered data. 
Computing blocks in the library are designed for applications
such as interferometry, pulsar dedispersion and timing, and transient
search pipelines. 
We describe the design and implementation of the Bifrost
framework and demonstrate its use as the backbone in
the correlation and beamforming back end of the Long Wavelength
Array station in the Sevilleta National Wildlife Refuge, NM.
\end{abstract}

\keywords{instrumentation: miscellaneous --- methods: data analysis --- methods: numerical --- methods: observational}

\ifarxiv
\else
  \clearpage
\fi

\tableofcontents

\ifclean

\else
  \fcolorbox{gray}{orange}{\texttt{\textbackslash todo}}
  \fcolorbox{gray}{blue!40}{\texttt{\textbackslash todostyle}}
  \fcolorbox{gray}{yellow!40}{\texttt{\textbackslash todofigure}}
  \fcolorbox{gray}{purple!40}{\texttt{\textbackslash todolatex}}
  \fcolorbox{gray}{green!40}{\texttt{\textbackslash tododetail}}
  \fcolorbox{gray}{olive}{\texttt{\textbackslash todotable}}
  \fcolorbox{gray}{teal}{\texttt{\textbackslash todocitation}}
  \fcolorbox{gray}{white}{\texttt{\textbackslash todoreply}}
  \listoftodos
\fi

\ifarxiv

\else
  \newpage
\fi

\section{Introduction}

Modern radio observatories are increasingly reliant on real-time processing
to overcome data storage and analysis bottlenecks. A trend toward
wider bandwidth digital systems, higher bitwidths of digitized signals,
and more numerous antennas has driven improvements in both the capabilities and data throughput
requirements of radio telescopes \citep{jones2012,hickish2016,price_spectrometers_2016}.

Consequently, astronomy survey data volumes are growing rapidly and are
increasingly burdensome to analyze \citep{berriman_how_2011}. As an example,
the Square Kilometer Array (SKA), scheduled to begin construction in 2018,
will produce on the order of 10 exabytes per day in buffer memory \cite{quinn_delivering_2015}
--- about $5\times$ the 2016 global internet traffic \cite{cisco_zettabyte_2016}.
The inability to store such volumes of data necessitates that they be processed
and analyzed in real-time, putting significant demands on computing hardware
and software.

As a result of computing demands, many-core accelerators --- specifically Graphics Processing
Units (GPUs) --- have become commonplace in both high-performance computing\footnote{As
of November 2016, 40 of the top 50 supercomputers on the Green 500 energy
efficiency ranking used accelerators} 
and radio telescope installations \cite[see][]{barsdell_analysing_2010},  such as
the Green Bank Telescope \citep[GBT, ][]{ransom_guppi:_2009};
the Canadian Hydrogen Intensity
Mapping Experiment \citep[CHIME, ][]{bandura_canadian_2014,recnik_efficient_2015};
the Low-Frequency Array \citep[LOFAR, ][]{serylak_observations_2013};
the HIPSR system at the CSIRO Parkes Observatory \cite{price_hipsr:_2016}; and
the Large-Aperture Experiment to Detect the Dark Ages \citep[LEDA, ][] {kocz_digital_2015}.

However, developing software for GPUs often requires expertise in massively-parallel programming,
knowledge of the memory hierarchy, and significant development effort.
The noted lack of software specialists working in astronomy (discussed
in \citet{lupton_sdss_2001}) complicates the development process more so.
In a time when the community is seeing impressive growth in the availability
of data, there is a lack of astronomers with an appropriate level of software
development experience. In this environment, the creation of a high-level, easy-to-use development
platforms is essential to ensuring that the community can develop
efficient processing pipelines to manage astronomical data.

In this article, we introduce an answer to this problem:
Bifrost, a high-throughput, GPU-enabled
stream-processing framework, created to expedite the development of high-performance
data processing pipelines in astronomy. This paper will present
the core concepts of Bifrost such as the blocks and pipelines abstraction
in \cref{sec:framework}, the technical details of Bifrost's implementation
in  \cref{sec:implementation}, and descriptions of included
Bifrost features and example pipelines in \cref{sec:included_modules}.
Pipelines made with Bifrost are compared performance-wise, and also regarding
usability, in \cref{sec:comparison_and_eval}.
Finally, the implementation
of Bifrost as the back end for the newest
Long Wavelength Array (LWA) station, LWA-SV
\cite[see][for the description of the
first LWA station, LWA1]{taylor_first_2012}, is outlined in \cref{sec:lwa}.

\subsection{Stream processing}

The Bifrost framework is specifically designed to process streaming data
in real-time; that is, to operate upon data with a potentially unlimited extent,
such as the time stream of digitized voltages from a telescope,
which has no clearly defined start time or end point. 
In Bifrost, streams of data are processed by creating pipelines:
a sequence of modular processing elements,
whose outputs feed into the inputs of the subsequent
elements. When an algorithm is pipelined, and
each processing element defines some finite minimum amount
of data to work on, the pipeline can be easily
set up for stream processing, by \textit{streaming}
the data through each element.

A real-time processing pipeline moves data through a series
of different filters, measurements, and transforms, which may reduce raw
data and flag for interesting occurrences such as bursts or other transients.
Traditionally, such processing pipelines are built from scratch, by manually
passing data between different systems. This development process
is inefficient, making pipelines time-consuming
to implement (demonstrated in \cref{sec:usability_vs_gpuspec}). For this reason,
processing frameworks have been adapted in some cases to ease the development
cycle for real-time processing pipelines by abstracting the data flow away
from the user.

\subsection{Existing packages}

The astronomy community has many software packages which encompass different
applications and use cases: the Astrophysics Source Code Library\footnote{\url{http://ascl.net/}}
\cite{allen_astrophysics_2013} now boasts more than 1500 codes, and has
seen the cumulative number of citations to its entries double every year since
2012 \cite[see][page 3]{allen_2016_2017} to over 1000 in 2016. The astronomy
community has a vast supply of codes, made
for specific tasks, but the process of integrating these algorithms
together into an efficient real-time pipeline remains
a difficult problem, and there is a lack of a standard stream-processing framework
for astronomical applications. Numerical computing libraries such as
\texttt{numpy}\footnote{\url{https://www.numpy.org}}
\citep[see][]{walt_numpy_2011}
provide high productivity, but lack the
performance required for use in cutting-edge applications.

A few astronomy pipeline libraries have been developed,
including \textsc{Pelican} \cite{mort_pelican:_2015}, \textsc{PSRDADA}\footnote{\url{http://psrdada.sourceforge.net/}},
and \textsc{HASHPIPE}\footnote{\url{https://github.com/david-macmahon/hashpipe}}.
% Some text from benbarsdell/bifrost/wiki
\textsc{Pelican} provides a highly abstracted Application Program Interface (API)
but is designed for
static, quasi-real-time processing purposes. \textsc{PSRDADA} and \textsc{HASHPIPE}
are designed for high-throughput stream processing, but suffer from rigidity
and low-level C APIs.

There are also frameworks not specifically designed for astronomy,
such as \textsc{GStreamer}\footnote{\url{https://gstreamer.freedesktop.org/}},
which was used in the Laser Interferometer Gravitational-Wave Observatory (LIGO)
merger detection pipeline \cite{messick_analysis_2017}.
However, the use of frameworks specialized for other domains requires laborious porting
of astronomy algorithms, file formats, and data types (e.g., support for complex numbers,
low-bit integers, etc.).

Bifrost addresses these issues by combining a high-level pipeline interface in Python
with a high-performance GPU-enabled back end that contains data structures and algorithms
relevant to astronomy data processing. This combination results in a framework that is
both highly productive and ready for cutting edge high-throughput applications.
Bifrost also includes detailed documentation and example pipelines that are easily adapted
to new applications. The use of Bifrost in a real-world application (the digital back end
of LWA-SV, discussed in \cref{sec:lwa}) also led to many refinements and
enhancements of the framework.

\begin{figure}
%\begin{wrapfigure}[10]{r}{0.4\textwidth} %Uncomment for final draft
\begin{center}
\includegraphics[width=\mediumfiguresize\textwidth]{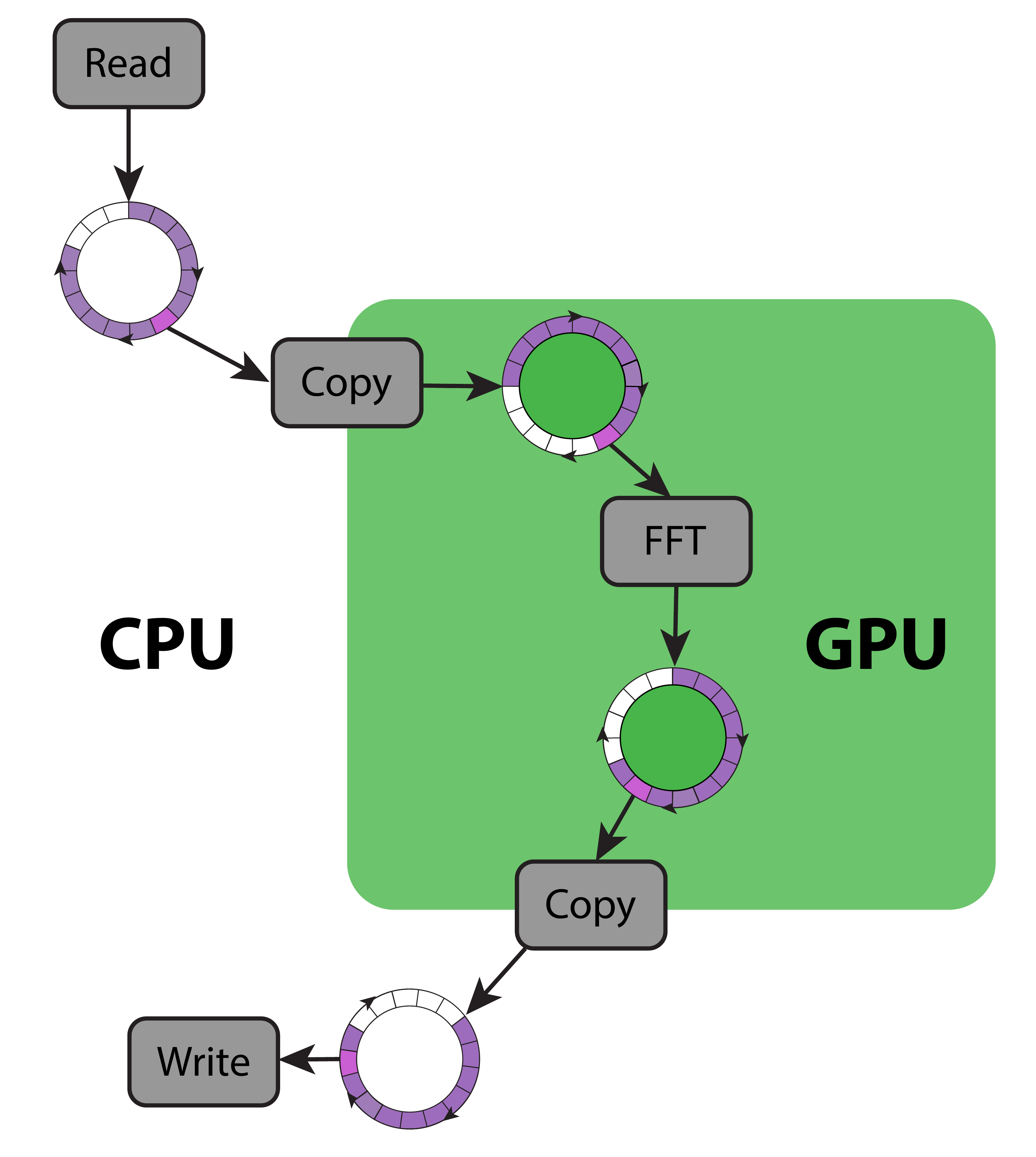}
\end{center}
\caption{A cartoon illustrating some of the basic concepts
of Bifrost: ring buffers (``rings''), blocks, and memory spaces.
Each block: ``Read'', ``FFT'' (a Fast Fourier Transform), and ``Write'',
lives in an independent
CPU thread and applies a user-defined operation
to a stream of data. Data are communicated from one block to
the next
through ring buffers --- contiguous allocations of memory that
emulate wrap-arounds from end to beginning.
Ring buffers can be written to and read from different threads
simultaneously, which enables
blocks to operate asynchronously.
}
\label{fft_pipeline:fig}
%\end{wrapfigure}
\end{figure}

\section{Overview of Bifrost}
\label{sec:framework}

Bifrost, currently versioned at v0.8.0,
combines a high-level pipeline interface with a collection of
high-performance data structures and algorithms relevant to astronomy
data processing. The framework -- written in \cpp and Python, with GPU
support enabled through NVIDIA's CUDA platform -- is divided into
\textit{front end} and \textit{back end} components, which are described
in the next two sections.

\subsection{Front end}
\label{sec:front_end}

Bifrost's front end is written as a set of Python modules which provide
a high-level interface to the library. The design uses multiple layers of
abstraction (all in Python):

\begin{enumerate}[leftmargin=5.5mm]
\item Block/pipeline abstractions,
\item Pythonic\footnote{We use \textit{pythonic} here to mean code that takes full advantage of the features of the Python language rather than being one-to-one mappings of the C API.} wrappers of back end functionality, and
\item Direct wrappers of the back end C library.
\end{enumerate}

At the highest level, the block/pipeline abstraction provides a
concise and intuitive means of developing processing pipelines.
Algorithms are described as blocks whose inputs and outputs can be
connected to form pipelines. Use of rich metadata enables blocks to
interoperate seamlessly. The block/pipeline abstraction is described
in more detail in \cref{sec:blockpipeline}.

Beneath the block/pipeline abstraction sits a set of Pythonic wrappers
around back end functionality. These provide a robust and convenient
interface to the key components of Bifrost's back end, such as its
GPU-enabled ring buffer (a data structure emulating a wrap-around in memory),
N-dimensional-array handling, and data
processing algorithms. In addition to supporting the block/pipeline
abstraction, this interface is used to implement a large suite of
unit tests. An abstract graphic of this block/pipeline/ring buffer
approach is shown in \cref{fft_pipeline:fig}.

Finally, Bifrost's back end C library is made directly accessible to
Python via a \texttt{ctypes}-based wrapper orchestrated through the
third-party \texttt{ctypesgen}
package\footnote{\url{https://github.com/davidjamesca/ctypesgen}}.
This provides comprehensive type-safe access to the C library, with an
interface generated directly from the library headers.

\subsection{Back end}
\label{sec:backend}

Bifrost's back end is written in \cpp and exposes a C API. It is
implemented as a collection of independent modules sharing a common
core of type definitions and utility functions. The key modules
(described in more detail in \cref{sec:included_modules}) are:

\begin{itemize}[leftmargin=5.5mm]
\item Ring buffer -- data structure for communication between threads in a pipeline.
\item FFT -- Fourier transform over any axes of an N-dimensional array.
\item Linear algebra -- matrix multiplication and related routines.
\item FDMT -- Fast Dispersion Measure Transform
\citep[the efficient dedispersion algorithm introduced in][]{zackay_accurate_2014}.
\item Quantize/unpack -- conversions to/from 1-32 bit data types.
\item Transpose -- efficient data reordering.
\item Map -- application of arbitrary functions to N-dimensional arrays.
\item User Datagram Protocol (UDP) capture/transmit -- receive/send data across a network.
\end{itemize}

The back end API uses a single structure, \texttt{BFarray}, to
represent N-dimensional arrays, providing a concise and consistent
interface to many routines. The structure describes the type and shape
of the data, along with other information such as the memory space in
which the data are stored (e.g., system memory, CUDA device memory).

\subsection{Blocks and pipelines}
\label{sec:blockpipeline}

Bifrost's high-level interface allows applications to be described as
sets of \textit{blocks} connected to form pipelines. Conceptually,
blocks are black boxes that process streams of data, with inputs and
outputs that can be connected. Blocks with no inputs or
outputs are referred to as \textit{source} and \textit{sink} blocks
respectively, while those with both are referred to as
\textit{transform} blocks.

The flow of data through a pipeline is logically partitioned using the
concepts of \textit{sequences} and \textit{spans}.

\subsubsection{Sequences}
Sequences define the boundaries between logically-independent
intervals of data flow; e.g., when processing a set of files or
observations, each file or observation is typically treated as a separate
sequence. Sequences each have their own \textit{header}, a collection of
metadata describing the contents of the sequence.
%Used to be "chunk"

\subsubsection{Spans}
Spans define the intervals of data within a sequence as they are
processed by a block. Blocks ``acquire'' (read) or ``reserve'' (write) spans
of data, process them, and then continue to the next span until the
sequence is complete. Importantly, blocks may request spans of any
extent and may progress through the sequence at any rate, with complete
independence from other blocks in the pipeline.

\subsubsection{Frames}
Within a span, Bifrost defines a \textit{frame} as a single element of
the data stream, analogous to the frames of a video stream. Frames are
treated as N-dimensional arrays, defined by a data type and
shape. Other metadata can also be included, such as axis labels,
scales, and units, which are used to provide seamless interoperability
between blocks.

\section{Details of Bifrost}
\label{sec:implementation}
This section provides a look at how Bifrost
implements the features and concepts
discussed in \cref{sec:framework}.
\Cref{sec:api} outlines the API for Bifrost usage, and
gives an overview of how to create a block within the Bifrost
API.
\cref{sec:headers} describes how metadata is used in the framework,
and \cref{sec:concurrency} discusses how parallelism is achieved
on both the CPU and GPU.
Finally, \cref{sec:ring_buffer_memory} details how different memory
spaces are used and how the Bifrost ring buffer is implemented.

\begin{comment}
\subsection{Python API}
\label{sec:api}

Bifrost is implemented using a high-level Python API, which works on 
top of a \cpp back end. The API 
can be separated into two families of
syntax---1. the block-creation syntax, which
allows the user to build a block in the context
of Bifrost, specifying how an algorithm reads
data and writes data to rings, and interacts with 
any outside processes, and 2. the
pipeline-creation syntax, which allows the user
to connect these blocks together, and run the 
final product. 

Bifrost has main dependencies on \texttt{numpy} and 
\texttt{ctypesgen}. A GPU-enabled version 
of Bifrost also depends on NVIDIA CUDA\footnote{\url{http://www.nvidia.com/object/cuda_home_new.html}}. 
Bifrost has a \cpp back end to enable efficient 
program execution and detailed memory 
management, and a high-level Python API 
to enable easy interfacing and pipeline 
generation. 
\end{comment}

\subsection{Block and pipeline APIs}
\label{sec:api}

\begin{figure}[H]
%\singlespacing
\begin{center}
\begin{minted}[mathescape,
               %linenos,
               numbersep=0pt,
               %gobble=2,
               framesep=2mm,
               bgcolor=codegray,
               fontsize=\scriptsize]{python}
from copy import deepcopy

import bifrost as bf
import bifrost.blocks as blocks
import bifrost.views as views
import bifrost.pipeline as bfp
from bifrost.ndarray import copy_array


class CopyBlock(bfp.TransformBlock):#                                         $\tikzmark{block-start}$
    """Copy the input ring to output ring"""
    def __init__(self, iring, space=None, *args, **kwargs):
        super(CopyBlock, self).__init__(iring, *args, **kwargs)
        if space is None:
            space = self.iring.space
        self.orings = [self.create_ring(space=space)]
    def on_sequence(self, iseq):
        ohdr = deepcopy(iseq.header)
        return ohdr
    def on_data(self, ispan, ospan):
        copy_array(ospan.data, ispan.data)#$\tikzmark{block-end}$


raw_data = blocks.read_wav(['audio_file.wav'], gulp_nframe=4096) 

gpu_raw_data = CopyBlock(raw_data, space='cuda')#                                $\tikzmark{gpu-start}$
chunked_data = views.split_axis(gpu_raw_data, 'time', 256, label='fine_time')
fft_output   = blocks.fft(chunked_data, axes='fine_time', axis_labels='freq')
squared      = blocks.detect(fft_output, mode='scalar')
transposed   = blocks.transpose(squared, ['time', 'pol', 'freq'])#$\tikzmark{gpu-end}$

host_transposed = blocks.copy(transposed, space='cuda_host') 
quantized       = blocks.quantize(host_transposed, 'i8')
blocks.write_sigproc(quantized)

pipeline = bf.get_default_pipeline()# $\tikzmark{pipeline-start}$
pipeline.shutdown_on_signals() 
pipeline.run()#$\tikzmark{pipeline-end}$


\end{minted}
\end{center}
\AddNote{block-start}{block-end}{block-start}{Bifrost block definition}
\AddNote{gpu-start}{gpu-end}{gpu-start}{GPU processing part of pipeline}
\AddNote{pipeline-start}{pipeline-end}{pipeline-start}{Start the pipeline}

\caption{An example of Bifrost Python syntax
which creates a new block to copy memory
from the GPU$\leftrightarrow$CPU,
and then uses to implement
a pipeline which
channelizes an audio file on the GPU,
writing it to a filterbank file.
}
\label{full-syntax:fig}
\end{figure}
\todofigure{I like this code sample, but I wonder if CopyBlock might not
be the best example of a block implementation, due to it needing to create
its own oring and it having a trivial on\_sequence (both of these things are unique
to CopyBlock).}

\subsubsection{Pipeline generation and static visualization}
\label{sec:static_vis}

In the Bifrost Python API, the \texttt{Pipeline} class
is used to set up a processing pipeline from 
a list of blocks and their connections. Blocks 
are initialized before the pipeline is
executed. This prior initialization allows
the user to store settings in the blocks before
they are executed by the pipeline. The pipeline object
then reads through all of the connections and
initializes ring buffers between each
connection with the appropriate memory space
and allocation.

\begin{figure}
{
\centering
\includegraphics[width=\fullfiguresize\textwidth]{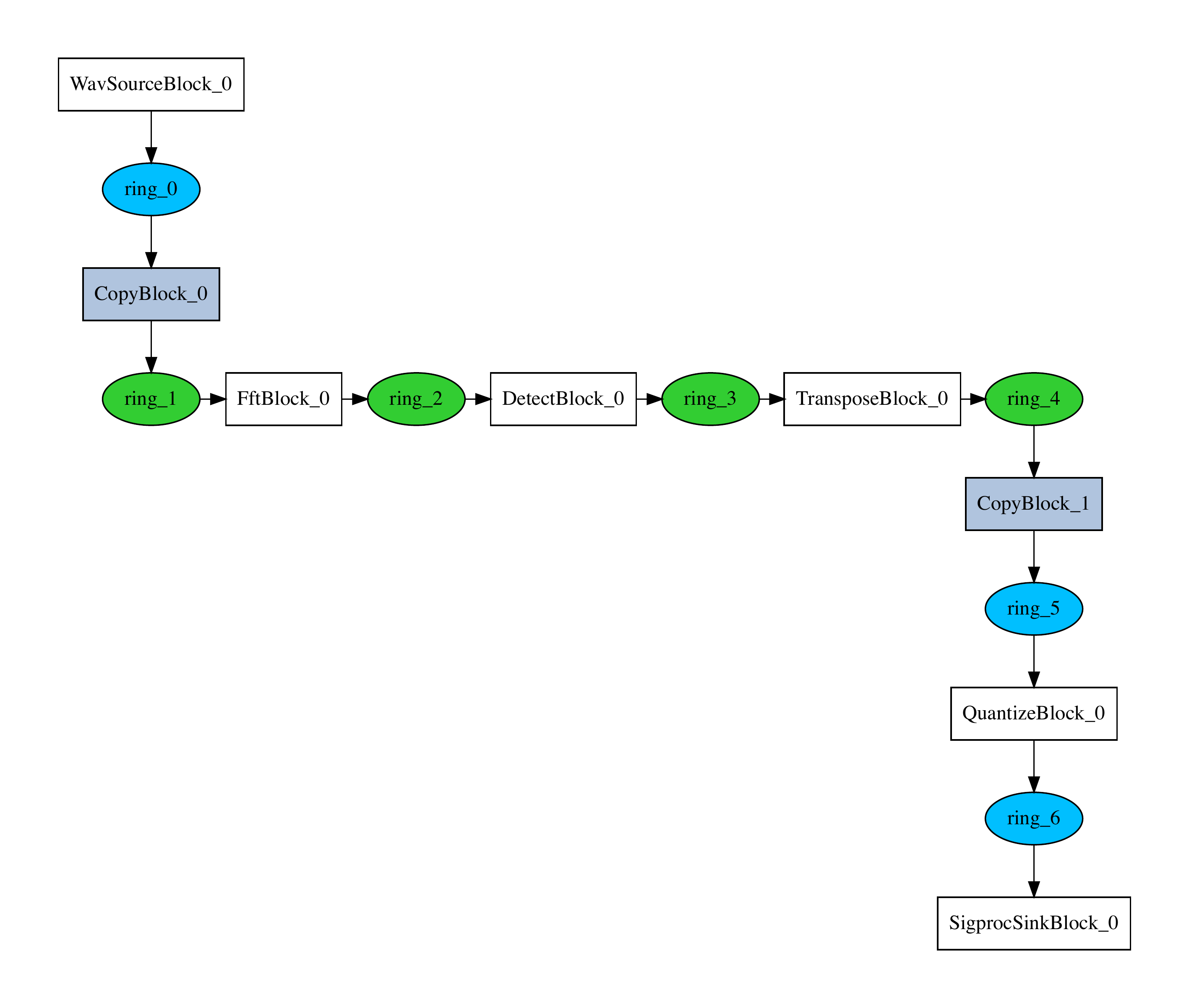}
\caption[Caption for LOF]{A graph generated by 
Bifrost for an example pipeline
created in the Python API, the same
one given in \cref{full-syntax:fig}.
Green color represents CUDA memory allocation,
and blue color represents system allocation.
This visualization 
uses Graphviz \cite{ellson_graphviz_2001},
which is wrapped by pyGraphviz\textsuperscript{\textdagger}.
}
\label{graph:fig}
}
\small\textsuperscript{\textdagger}\url{https://github.com/pyGraphviz/pygraphviz}
%\end{wrapfigure}%
\end{figure}

After the pipeline class performs this,
the user can trigger a static visualization
function: \texttt{Pipeline.dot\_graph()}, which generates a graph of the pipeline
using Graphviz. An example of this is shown
in \cref{graph:fig}.

Stating \texttt{Pipeline.run()} executes the pipeline:
the pipeline will create a thread for each block,
which run until their blocks finish
processing, until the user performs a keyboard
interruption, or until a custom signal defined
by \texttt{Pipeline.shutdown\_on\_signals()} is activated.

\subsubsection{Views}
In addition to blocks, Bifrost also defines \textit{views}, which
allow for simple metadata transformations that require no actual
modification of the data stream. Examples include the \texttt{astype}
view, which allows data to be explicitly reinterpreted with a different data
type (analogous to \texttt{reinterpret\_cast} in \cpp or
\texttt{numpy.ndarray.view} in Python),
and the \texttt{split\_axes} and \texttt{merge\_axes} views,
which allow modifying the shape of data frames.

\subsubsection{Block scopes}
\label{sec:block_scopes}
Bifrost's high-level pipeline/block API uses a hierarchical design
that allows blocks to be nested inside \textit{block scopes}. These
constructs encapsulate parameters that are inherited by the blocks
within; inherited parameters include the CPU core and GPU to which the
blocks are bound. Block scopes also expose a \textit{fuse} flag that
configures blocks to share input, output, and temporary buffers; this
can significantly reduce the amount of memory required to execute a
pipeline, at the cost of the blocks being unable to execute in
parallel.

An example of \texttt{block\_scope} usage is given in
\cref{fig:block_scope}.

\begin{figure}[h]
\begin{center}
\begin{minted}[mathescape,
               %linenos,
               numbersep=5pt,
               %gobble=2,
               bgcolor=codegray,
               framesep=2mm,
               fontsize=\scriptsize]{python}
import bifrost as bf
...
with bfp.block_scope(core=2, gpu=0, fuse=True):
    data = bf.blocks.transpose(data, ['pol', 'freq', 'time']
    data = bf.blocks.fdmt(data, max_dm=1000.)
    data = bf.blocks.transpose(data, ['time', 'pol', 'dispersion'])
...
\end{minted}
\end{center}
\caption{Code snippet which shows the use of a \texttt{block\_scope}
in a dedispersion pipeline. The three blocks inherit the
\texttt{core} and \texttt{gpu} parameters such that they will
all be executed on the same CPU core and GPU. Because the \texttt{fuse}
flag is enabled, they will also share input and output buffers,
which reduces memory usage but forces them to execute sequentially.}
\label{fig:block_scope}
\end{figure}

\subsubsection{Guarantees}
\label{sec:guarantees}
By default, blocks are \textit{guaranteed}, which means that the block
that feeds them will (if necessary) wait for them to finish reading
before writing new data. This results in blocks applying back-pressure
to the pipeline if their execution is too slow. In some cases it is
desirable to prevent this by making blocks \textit{unguaranteed}; the
downside is that their input data will then be ungraciously
overwritten if they are too slow, and they will end up
\textit{skipping} data. This data-guarantee functionality
can be useful for
diagnostic or monitoring blocks, or for decoupling different branches of
a pipeline.

\subsubsection{Block definitions}
\label{sec:block_defn}
A block implementation must
define two functions: \texttt{on\_sequence}
and \texttt{on\_data}. The \texttt{on\_sequence} function is called at
the beginning of each sequence and is where the block interprets the
input header(s) and generates its output header(s). The
\texttt{on\_data} function is called whenever a new set of input and
output spans have been acquired or reserved and is where the block
performs its processing operation. While more complex data processing
patterns are possible, this simple abstraction is sufficient for most
blocks, and greatly simplifies the process of implementing new blocks.

A block can also, optionally, redefine the function
\texttt{on\_skip}, which defines
the behavior of the block when data is skipped in a ring
if it is set to \texttt{guarantee=False} (see \cref{sec:guarantees}).
By default, the output data is set to zero.

\subsection{Metadata headers}
\label{sec:headers}

Bifrost makes extensive use of metadata stored in sequence headers.
These metadata give semantic meaning to the contents of a ring, and
enable blocks to communicate seamlessly with little or no user
intervention.
While sequence headers may be defined any way the 
user wishes, so long as they are consistent enough
to write blocks which can interpret those headers,
Bifrost has a ``standard'' for headers.
Abiding by this (informal) specification allows the user to make use
of Bifrost's pre-defined blocks with ease. A list
and descriptions of the built-in Bifrost blocks are given
in \cref{sec:included_modules}.

Bifrost's standard header is as follows, which
is inspired by several of the attribute names used for
\texttt{numpy} arrays.

\begin{itemize}[leftmargin=5.5mm]
\item \texttt{`name'} - Name the sequence. This could be the
filename that was used for data input, for instance.
\item \texttt{`\_tensor'} - Dictionary of the following keys:
\begin{itemize}[leftmargin=5.5mm]
    \item \texttt{`dtype'} - Datatype of the data, in Bifrost format,
    e.g., \texttt{`i8'} for 8-bit signed integers, or
    \texttt{`cf32'} for 32-bit
    complex floating point numbers.
    \item \texttt{`shape'} - The shape of each frame of data. E.g.,
    for each time step of a time stream, this would indicate a dimension
        of the frequency channels, and perhaps another dimension for
        polarization. The axis which is ``streamed'', such as
        time, should have a value of -1.
    \item \texttt{`labels'} - List of labels for each of the axes in shape, in
    the same order. Some blocks may assume a specific name for a label; others
    will work on whatever axis you specify.
    \item \texttt{`units'} - List of physical units each axis of a stream.
    For a time stream, the axis with -1 size (the
    streamed axis), could have units of \texttt{`s'} for seconds. Units
    are made possible using the \texttt{pint} library\footnote{\url{https://github.com/hgrecco/pint}}.
    \item \texttt{`scales'} - List of
    2-tuples which specify the offset and spacing between values along
    each axis (specified in the same order). The first element of
    a tuple is the starting offset of the axis, such as the starting
    Unix time of a time stream, while the second element
    is the spacing between elements along the axis, in units of the
    corresponding `units' entry.
\end{itemize}
\end{itemize}

Any other dictionary values can be added after this, to specify
observational parameters or anything else that is relevant
to blocks downstream in the pipeline.

\Cref{fig:header} shows an example header for the \texttt{.wav}-reading Block:

\begin{figure}[h]
%\singlespacing
\begin{center}
\begin{minted}[mathescape=true,
               %linenos,
               numbersep=5pt,
               %gobble=2,
               framesep=2mm,
               bgcolor=codegray,
               fontsize=\scriptsize]{python}


{
    'frame_rate': 44100,
    'gulp_nframe': 4096,
    '_tensor': {
        'units': ['s', None],
        'dtype': 'i16',
        'shape': [-1, 2],
        'labels': ['time', 'pol'],
        'scales': [[0, 2.2675e-05], None]
    },
    'name': 'my_music.wav',
    'time_tag': 0
}

\end{minted}
\end{center}
\caption{
Example header for a Bifrost sequence, which is in this case,
output from \texttt{bifrost.blocks.read\_wav}.
Since a ring buffer in Bifrost does not have
a particular data type or shape or size, the
header for each sequence on that ring buffer
must provide all information necessary to interpret the
incoming stream of bytes.
The header is read before the data is processed,
which lets the block store various settings
and options targetted by the information given in the header.
Headers should describe at the very least, the
data type and shape.
}
\label{fig:header}
\end{figure}

\subsection{Block execution and concurrency}
\label{sec:concurrency}

Bifrost involves multiple types of
parallelization to take advantage 
of all processors available. The first
level of parallelization - across different 
CPU cores - is handled by the threads, 
and by native Bifrost core-binding 
control. Each block executes in a separate thread.
The second level of 
parallelization, which is used for 
GPU processing, works by calling 
CUDA kernels from within a Bifrost block,
accessed via \texttt{ctypes}.

\subsubsection{\texttt{ctypes} and the Python Global Interpreter Lock (GIL)}
\label{sec:gil}
Most Python implementations (e.g., the default,
most commonly used \texttt{CPython}) are interpreters,
meaning that the user's Python 
installation has an executable 
(the Python interpreter) which 
reads through the Python code at
execution time, and executes 
machine instructions directly 
based on that executable's interpretation 
of each line of code. The GIL
of Python is a lock on the 
interpreter so that it can not
execute two lines of code simultaneously. 
This means that parallelizing Python code
is impossible unless the GIL is released. 

This means that Bifrost must evoke external
functionality to escape the GIL and
execute truly multi-threaded pipelines.

\texttt{ctypes} is a native library for Python which allows
the interpreter to execute arbitrary C functions from a
dynamically loaded shared library. \texttt{ctypes} releases the GIL while
the C function is executing, allowing it to run with true concurrency.
Python’s \texttt{numpy} library is
an epitomic example of \texttt{ctypes}: most \texttt{numpy} functions release
the GIL, so calling these functions allows parallelization within
a pipeline.

\texttt{ctypes} wrappers for Bifrost's back-end are currently generated
using the third-party
\texttt{ctypesgen}\footnote{\url{https://github.com/davidjamesca/ctypesgen}}
library, which parses Bifrost's C
headers and outputs Python code when the library is built.

\subsubsection{CPU parallelization}
\label{sec:cpu_parallelization}

\begin{figure}[H]
\begin{center}
\includegraphics[width=\fullfiguresize\textwidth]
{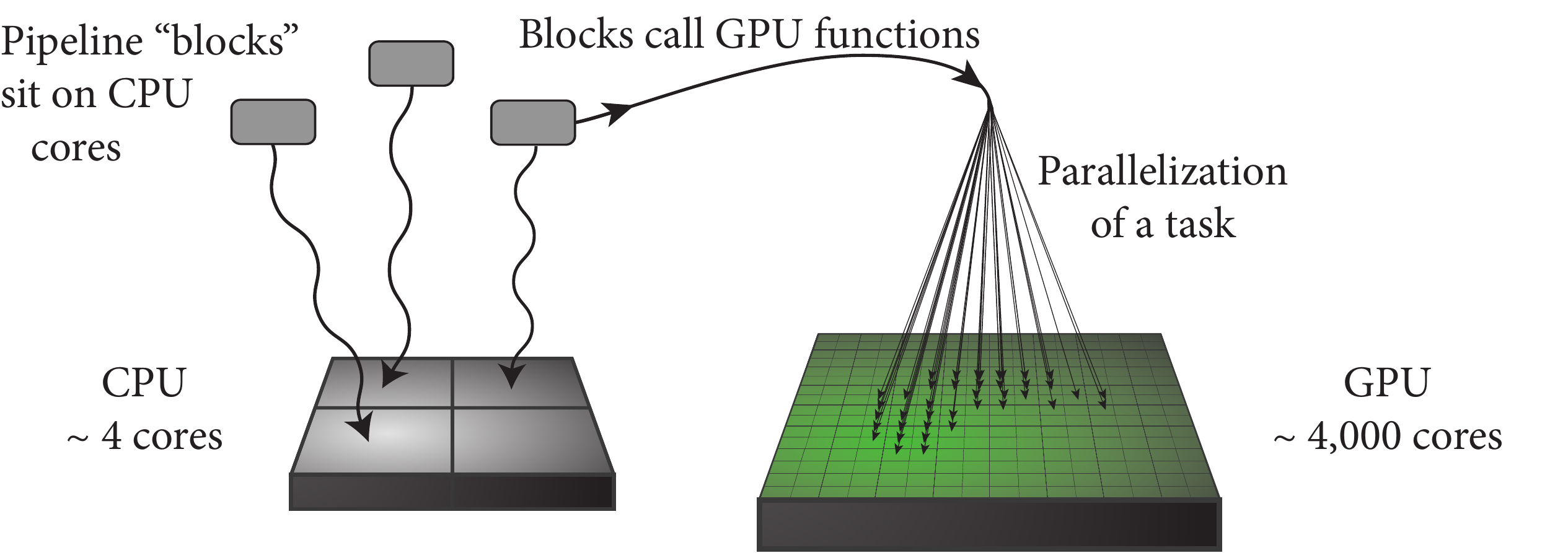}
\end{center}
\caption{A cartoon illustrating
the two types of parallelization
intrinsic to Bifrost---CPU, which
maps process blocks to different
CPU cores, and GPU, which 
allow CUDA functions to be called
on ring buffers.}
\label{syntax:fig}
\end{figure}

The first level of parallelization is done across CPU
cores. Python, where Bifrost pipelines are generated and
controlled from, has a native module called
\texttt{threading}. This module allows Bifrost to
put each block into a separate thread, so
that they can call algorithms without requiring these calls to
be sequential in code. This is represented visually
in \cref{syntax:fig}. As discussed in
\cref{sec:gil}, the use of
\texttt{ctypes} allows Bifrost to avoid Python’s
GIL, so that each block can be bound
to a specific CPU core, and process data at the same time.
An entire section of a pipeline can be bound to
a CPU core using a block scope (see \cref{sec:block_scopes}).

\subsubsection{GPU parallelization}
\label{sec:gpu_parallelization}

\subsection{Memory management and ring buffers}
\label{sec:ring_buffer_memory}

\begin{figure}[H]
\begin{center}
%Left bottom right top
\includegraphics[trim=50 0 0 0,clip,width=\fullfiguresize\textwidth]{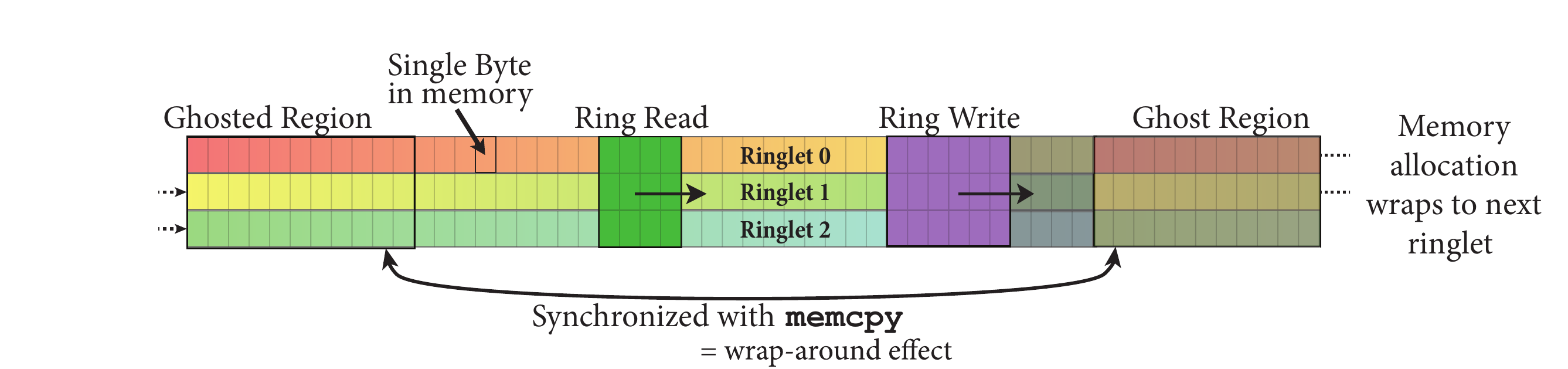}
\caption{Bifrost ring buffer implementation, showing ringlet sub-structure.
Each ringlet is contiguous in memory and follows the previous
ringlet. The ghosted and ghost regions are identical in memory content,
and are synchronized with \texttt{memcpy}, so that a 
ring reader or writer can always read contiguous chunks
of memory, even if its gulp size is not a divisor
of the total ring size (and so partially overlaps
the ghost region at the end of the ring buffer). The
gray color in the diagram represents memory which has
not been written to yet in this example.
}
\label{ring:fig}
\end{center}
\end{figure}

A ring buffer (also called a circular buffer)
is a data structure created to emulate a 
wrap-around in memory. 
A diagram of the ring buffer in Bifrost is given in
\cref{ring:fig}.
A ring buffer is a contiguous
block of memory that 
pretends as though it has an
arbitrary beginning and no end. When two processes are 
working on a contiguous block of memory that is not
circular, with one process writing 
and the other reading, the writing process must either
stop at the end of the block, 
allocate more memory, or go back to the start
of the block of memory. With a ring
buffer, the writing process wraps around to the start
of the block of memory and
continues to write seamlessly. This wrap-around is
emulated with a region of memory called
the “ghost region,” which is an extra allocation
of memory at the end of a ring 
buffer that is synchronized with the beginning of the ring
buffer. This circular interaction
gives the term its name. Ring buffers have been
used successfully as a standard data
structure for buffer management in other
astronomy stream processing software in
the past, such as
PSRDADA\footnote{\url{http://psrdada.sourceforge.net/}},
\textsc{Pelican}
\cite{mort_pelican:_2015},
\texttt{kotekan} \cite{recnik_efficient_2015},
and
GUPPI \cite{ransom_guppi:_2009}, which has
evolved into the more general
\textsc{HASHPIPE}\footnote{\url{https://github.com/david-macmahon/hashpipe}}.

In Bifrost, ring buffers form the 
connections between blocks in a 
directed graph. A ring has a single
block that writes to it and can 
be read from multiple blocks at 
the same time. In addition to this,
Bifrost ring buffers are dynamic: 
they can be resized by the user 
during processing
to allocate more memory 
for storage.  

The \textit{gulp size} of a ring buffer refers
to how many bytes of memory a block should read
in from the ring buffer at once.
The \textit{buffer factor} of a ring buffer relates
to the size of the ring buffer in memory
as a multiplicative factor of the
\textit{gulp size}.
By default, the buffer size is 4, which means
that 4 times the \texttt{gulp\_size} is the total
memory of a ring buffer. As the gulp\_size changes
with a new incoming sequence, the ring buffer
will resize accordingly. Different buffer sizes
can also be set by the user.

Block fusion, which is discussed in \cref{sec:block_scopes},
uses a buffer size of 1 for the rings between blocks,
allowing things to operate sequentially. This is useful
when memory is scarce, especially for
GPU-only parts of a pipeline.

Bifrost ring buffers can be bound to
one of the following memory spaces:
\begin{itemize}
\item \texttt{system} --- host memory allocated with \texttt{aligned\_alloc}
\item \texttt{cuda\_host} --- pinned host memory
    allocated with \texttt{cudaHostAlloc}
\item \texttt{cuda} --- device memory allocated with \texttt{cudaMalloc}
\item \texttt{cuda\_managed} --- unified memory allocated
with \texttt{cudaMallocManaged}
\end{itemize}

Ring buffers store 
their memory binding in the “space” 
attribute: when a Bifrost block 
reads in a ring buffer, they can
check this attribute to trigger 
CPU or GPU processing depending
on the space of the ring buffer.
Bifrost blocks can then call the
appropriate C functions on that 
data using \texttt{ctypes}.

Copying data from host memory to GPU
memory is as easy as inserting
a copy block into a pipeline, specifying the
desired output space, such
as \python|g_data = copy(h_data, space='cuda')|.
These memory spaces are shown graphically in \cref{fft_pipeline:fig},
and also in \cref{syntax:fig}.

\subsubsection{Sequences and spans}
Bifrost ring buffers are implemented as contiguous memory allocations.
Users of a ring buffer can request that it be resized to satisfy
a given minimum total size as well as a maximum guaranteed-contiguous
span. Blocks use this mechanism to specify the gulp size in which they
will access memory in the ring.

Sequence metadata, including their starting and (eventually) ending
offset, name, time tag and header, are stored with the ring.
Reading is always done relative to the beginning of a sequence, and is
bound by the beginning and end of the sequence (attempts to read outside
the bound of a sequence will result in either no data or an error,
depending on context).

\subsubsection{Ringlets}
\label{sec:ringlets}
To support data streams in which the temporal dimension is the
fastest-changing dimension in memory, Bifrost rings use a concept dubbed
``ringlets''. Ringlets act like a second dimension on a ring, emulating
multiple individual rings running parallel to each other. The ability to
store data in a time-fastest manner is important for the efficient execution
of some algorithms. Moving data from time-slowest to/from time-fastest
requires a transpose operation (the transpose block is a common feature of
data-processing pipelines).

\subsubsection{Random access}
A common use-case in real-time data processing pipelines is to perform
triggered dumps of buffered data. Bifrost ring buffers support
random access to sequences by name or time tag which allows
the user to easily inspect and save
a specific portion of data that are still in memory.

\section{Bifrost libraries and tools}
\label{sec:included_modules}
\subsection{Some included blocks}
\label{sec:included_blocks}

Bifrost currently has the following built-in blocks:

\begin{itemize}
\item \texttt{accumulate} - Accumulate frames one at a time.
\item \texttt{copy} - Copy data, possibly between memory spaces (e.g.,
CPU $\rightarrow$ GPU).
\item \texttt{detect} - Apply square-law detection to I/Q data to form
polarization products.
\item \texttt{fdmt} - Fast Dispersion Measure Transform (see \citet{zackay_accurate_2014}).
\item \texttt{fft} - Fast Fourier Transform, using the CUFFT
library\footnote{\url{https://developer.nvidia.com/cufft}}.
\item \texttt{fftshift} - Shift the frequency center of an array along its
axes.
\item \texttt{print\_header} - Print the
header of each new incoming sequence to stdout.
\item \texttt{transpose} - Reoder the axes of N-dimensional data.
\item \texttt{read\_sigproc}/\texttt{write\_sigproc} - Read or write
data in \textsc{SIGPROC} files\footnote{\url{http://sigproc.sourceforge.net/}}.
\item \texttt{read\_wav}/\texttt{write\_wav} - Read or write data in 
Wave files.
\item \texttt{read\_audio} - Read data from an audio stream using the
\texttt{portaudio} library\footnote{\url{http://www.portaudio.com}}.
\item \texttt{read\_guppi} - Read data from GUPPI
RAW\footnote{\url{https://github.com/UCBerkeleySETI/breakthrough/blob/master/doc/RAW-File-Format.md}} files.
\item \texttt{serialize} - Write data to a Bifrost-specific
file structure, with headers as JavaScript Object Notation (JSON),
and data streams as raw binary.
\item \texttt{quantize} - Quantize data to a given smaller data type.
\item \texttt{unpack} - Unpack packed data to a given larger data type.
\item \texttt{reverse} - Reverse data along a specific axis.
\end{itemize}

\subsection{Some included views}
\label{sec:included_views}
Bifrost views are all built upon a single
function, \texttt{block\_view}, which takes a block
and a ``header transform,'' which is a function
acting on a header, and applies that function to the block
to modify the outgoing sequence headers.
These views are easy to create, but Bifrost has a few
built-in ones to speed up development:

\begin{itemize}
\item \texttt{rename\_axis} - Change the label string on a specific
axis.
\item \texttt{split\_axis} - Split an axis into chunks to form a new axis.
\item \texttt{merge\_axes} - The inverse of split\_axis: merge two axes into one.
\item \texttt{astype} - Change the datatype without changing the underlying data.
\item \texttt{expand\_dims} - Add an extra dimension to the shape.
\end{itemize}

\subsection{The Bifrost ndarray implementation}
\label{sec:ndarray}
Much of Bifrost is built around a featureful multi-dimensional
array class, which inherits from \texttt{numpy.ndarray} for compatibility.
Bifrost's \texttt{ndarray} is the main part of the interface
between ring buffers and the functions which
compute on their data and
provides an easy
way to interface with both CUDA allocations
of memory and \texttt{numpy} arrays.

Bifrost's \texttt{ndarray} is cached with the
C-struct: \texttt{BFarray}, which supports
many datatypes and memory spaces, and is used
to link with all of the C/\cppnospace/CUDA back end of Bifrost.

\texttt{ndarray} also has a simple function,
\texttt{as\_GPUArray}, which produces a
\texttt{PyCUDA}\footnote{\url{https://github.com/inducer/pycuda}}
GPU allocation object, for easy interfacing with
\texttt{PyCUDA} functions \cite{klockner_pycuda_2012}.

\subsection{Bifrost {\tt  map}}
Bifrost's \texttt{map} routine allows arbitrary functions to be
applied to sets of ND-arrays using the GPU. The user provides a set of
named arrays along with a code string describing how their elements
are to be transformed, and the \texttt{map} routine generates and
executes a corresponding GPU kernel.

The following example demonstrates the use of \texttt{map} from
Python to multiply an array by 3:
\begin{minted}[mathescape,
               %linenos,
               numbersep=0pt,
               %gobble=2,
               framesep=2mm,
               bgcolor=codegray,
               fontsize=\small]{python}
bf.map('y = x * 3', {'x': x, 'y': y})
\end{minted}
In this example, \texttt{x} and \texttt{y} are \texttt{bf.ndarray}
objects whose data reside in GPU memory, and the operation is applied
to each element of the arrays. More complex operations can be performed
using explicit indexing, such as in this example that computes the
outer product of two arrays:
\begin{minted}[mathescape,
               %linenos,
               numbersep=0pt,
               %gobble=2,
               framesep=2mm,
               bgcolor=codegray,
               fontsize=\small]{python}
bf.map('c(i,j) = a(i) * b(j)', axis_names=('i', 'j'),
       data={'a': a, 'b': b, 'c': c})
\end{minted}
Other supported features include negative indexing, broadcasting,
support for scalar parameters, and complex math.

Internally, the \texttt{map} routine uses the
NVRTC\footnote{\url{http://docs.nvidia.com/cuda/nvrtc/index.html}}
library to just-in-time compile CUDA code into PTX, which is then
passed to the CUDA driver for execution. Compiled kernels are
cached in thread-local storage so that subsequent calls are fast.
The routine is part of the Bifrost back end, and can be called from
C or Python.

Due to its flexibility, \texttt{map} is used to implement some
blocks in the Bifrost library, including FFTShiftBlock, ReverseBlock,
DetectBlock, and AccumulateBlock. Using \texttt{map} avoids the need
to write custom kernels for each of these tasks, which is a
time-consuming and error-prone process.

\subsection{Real-time logging}
\label{sec:logging}
Bifrost implements lightweight real-time logging using
a {\tt proc}-like interface
(\texttt{procfs} is a special Linux filesystem
that lists system processes in a directory tree, and is used
for \texttt{ps} and \texttt{top}).  This
logging captures various aspects of each
block, e.g., CPU core binding and
timing information, and is useful not
only for debugging but also profiling
and monitoring running pipelines.  The
log files are written to the
Linux shared memory temporary file storage
in a hierarchical structure.  Similar
to {\tt proc}, each running pipeline
is a directory named by its
process ID. Within this pipeline
directory, there are sub-directories, one for
each block which implements logging. 
The log files are found within
these block directories and store information
in text files are \textit{key:value} pairs.
The log files are structured so
that each file describes a particular
aspect of the block.  For
example, the {\tt bind} files contain
how many and which CPUs or
GPUs are being used by a
block while the {\tt perf} stores
timing information on how long it
takes to load data from a
ring buffer.  This distribution
of logging information makes it possible
for the pipeline to update only
a subset of files when new
information is available and simplifies writing
software that aggregates and analyzes the
data.

\subsection{Tools}
\label{sec:tools}

Bifrost also comes with several pipeline-monitoring tools,
which can be used to expedite development with Bifrost. These
are found in the {\tt tools} folder in the Bifrost repo.
They make use of Bifrost's logging in Linux shared memory, i.e., {\tt /dev/shm}.  The tools included with Bifrost are:
\begin{itemize}
\item \texttt{like\_ps.py} - Shows which
blocks are running in a Bifrost pipeline, just like
\texttt{ps}.
\item \texttt{like\_top.py} - Provides a per-block analysis of the
processing load and timing of each of
the elements in a Bifrost pipeline
in quasi-realtime, just like {\tt top}.
\item \texttt{like\_bmon.py} - Provides information
on packet loss, which is useful for UDP packet capture
pipelines.
\item \texttt{pipeline2dot.py} - Creates graph files describing a running pipeline suitable for plotting.
\item \texttt{getirq.py} - Displays system interrupt core bindings.
\item \texttt{setirq.py} - Sets system interrupt core bindings.
\item \texttt{getsiblings.py} - Shows which cores are hyperthread siblings.
\end{itemize}

\section{Comparison and evaluation}
\label{sec:comparison_and_eval}

\begin{comment} %Until further notice.
\subsection{Comparisons with a serial program}

\tododetail{Create a comparison with 
some serial equivalent pipelines.}
\end{comment}

\subsection{Performance}
\label{sec:performance}

\begin{figure}[h]
\begin{center}
\includegraphics[width=\fullfiguresize\textwidth]{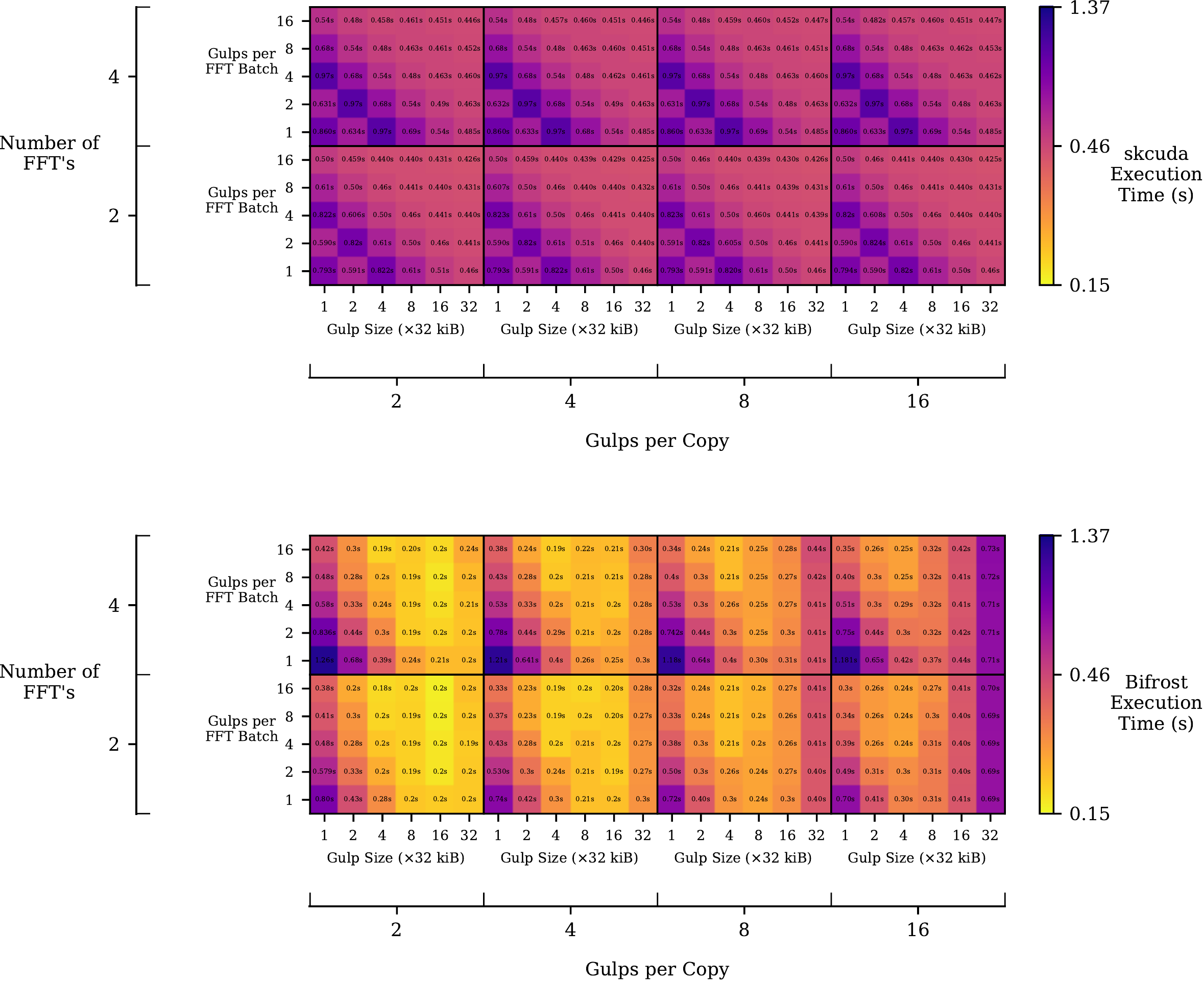}
\end{center}
\caption{Performance timings
of GPU pipelines for
Bifrost and scikit-cuda. Each
pixel is an individual benchmark.
The pipeline for each package
reads 32-bit floating
point data from a binary file, copies the
data to the GPU, then does a particular number of
Fast Fourier Transforms.
The data is read, copied, and transformed in
``gulps'' of a specific size.
Both pipelines are written in Python, have
no tuning of settings, and are set
to use the same file size,
gulp sizes, number of copies, and number
of batched FFTs. The times are
averaged over 56 trials,
and the significant figures reflect
standard deviation about the average
timing for each test.
Explicit combined errors are shown in \cref{fig:performance}.
The number inside each pixel represents the corresponding
measurement for each benchmark.
}
\label{fig:timings}
\end{figure}

\begin{figure}[h]
\begin{center}
\includegraphics[width=\fullfiguresize\textwidth]{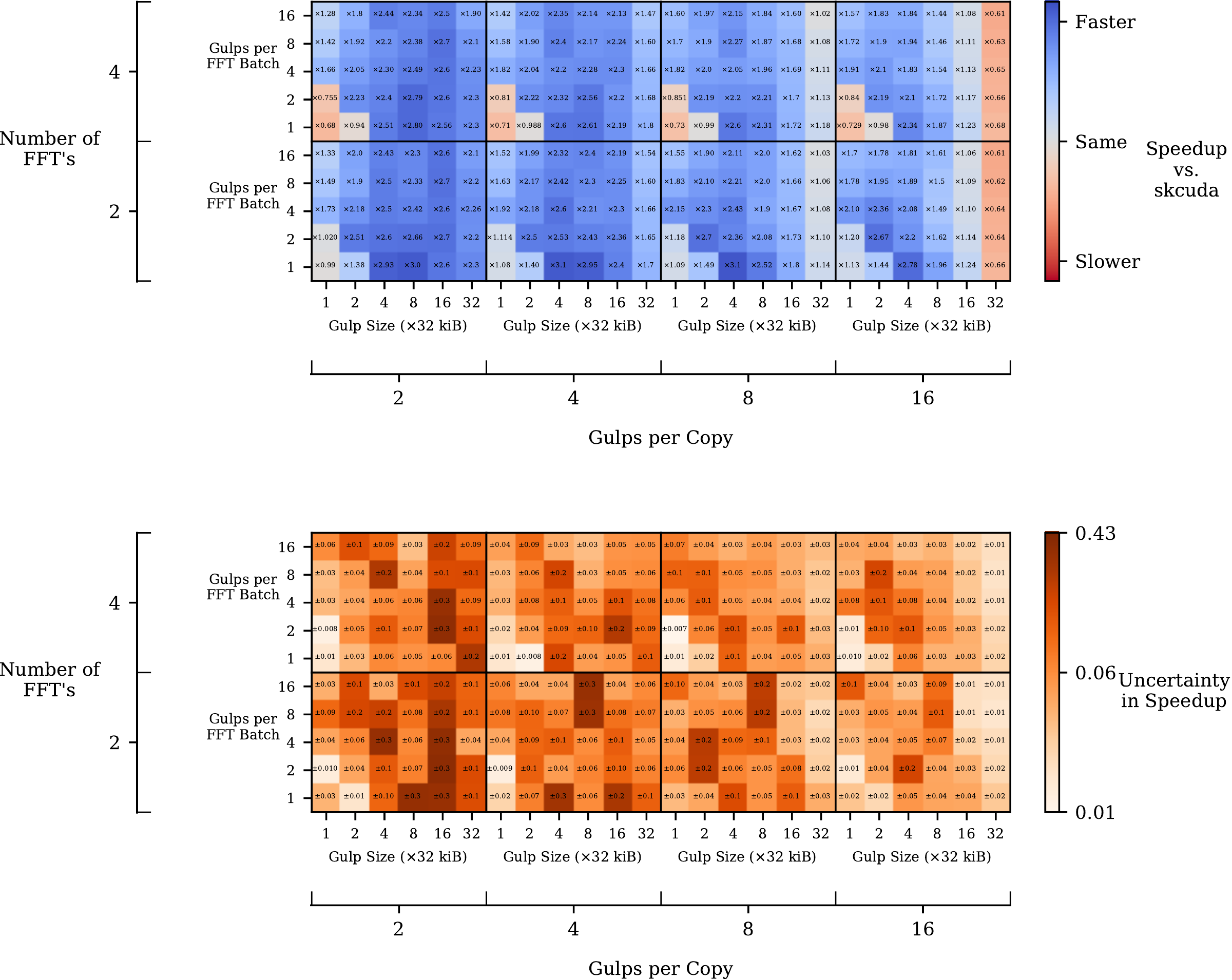}
\end{center}
\caption{Performance comparison
of GPU pipelines for
Bifrost and scikit-cuda,
using the timings shown in \cref{fig:timings}.
The bottom plot shows the combined error
from each pixel's standard deviation about
the average timing for
both of the plots in \cref{fig:timings}.
}
\label{fig:performance}
\end{figure}

In \cref{fig:timings,fig:performance}, the performance of
Bifrost is compared over various parameter
settings with a serial pipeline implemented
with the
package scikit-cuda\footnote{\url{https://github.com/lebedov/scikit-cuda}}
\cite{lev_givon_2015_40565}.
Both scikit-cuda and Bifrost
have Python frontends, wrapping a
cuFFT back end, meaning that differences
in performance should measure the
speedup due to pipeline parallelism,
or slowness due to Python overhead.

Due to the pipeline parallelism made available by
Bifrost, performance increases are seen in
\cref{fig:performance}, despite both packages
using the same back end cuFFT kernel.

In \cref{fig:timings}, a number of trends can be
seen in the lower figure (Bifrost).
As seen in each 5x6 pixel sub-grid, from left-to-right,
a greater gulp size results in significantly
increased performance of Bifrost, up to a certain point
where the GPU becomes saturated, and pipeline
parallelism loses effectiveness (the far right column
of each sub-grid). This is because smaller gulp sizes
result in more Python operations per kernel execution, and
slow down a pipeline. From bottom-to-top in each sub-grid,
more batched FFT's result in more efficient computation due to
extra parallelism. From the bottom row of sub-grids
to the top row, a slight performance decrease is seen, simply
because the GPU is performing twice as many FFT's.

In the top plot of \cref{fig:timings}, the
scikit-cuda pipeline shows more consistent, but
much slower values on the majority of benchmarks.
The consistency is due to the simpler operation of the library,
which acts as a thin wrapper on CUDA operations. This means there
will be fewer potential Python operations per kernel execution,
even on smaller gulp sizes. However, the relative slowness is due to
the lack of pipeline parallelism.

Then, in \cref{fig:performance},
this speedup is shown across the same grid of benchmarks,
highlighting the speedup Bifrost achieves when using
a modest gulp size, and demonstrating the magnitude
of speedup one can get on a GPU with pipeline parallelism.

\begin{comment}%TODO Waiting on this figure.
Using the NVIDIA tool \texttt{nvprof},
the pipeline parallelism available through Bifrost
is evident in the graphical representation
of CUDA streams (asynchronous memory copies)
being overlapped with GPU computation.
This graphical display is shown in \cref{fig:nvprof}.
\todofigure{Need to add this display with
the streams labeled.}
\end{comment}

\begin{comment} %Ben's computer:
on a computer with a quad-core
Intel Core i7-6700K CPU (4.2 GHz), with
16 GB RAM, and an EVGA GeForce GTX 1080 (SC
edition) GPU,
\end{comment}
The benchmarks described were measured
on an Amazon Web Services (AWS)
p2.xlarge instance, running Ubuntu 16.04,
inside a docker container using the
\texttt{nvidia/cuda}\footnote{\url{https://github.com/NVIDIA/nvidia-docker}}
base instance, running CUDA 8.0.
The p2.xlarge instances run a Tesla K80 GPU on an
Intel Xeon E5-2686v4 CPU, with 61 GiB of memory.
The benchmarks were performed using
Bifrost version \texttt{v0.8.0-benchmark},
with all optimizations turned on.
scikit-cuda was downloaded
at git commit hash b32a679.

\subsection{Developing with Bifrost}
\todo{Ben: This section reads OK, but it's not clear what its purpose is.}
Bifrost provides a variety of interfaces for working with
the library that help improve the productivity of writing
new software. The most obvious example of this
is the high-level Python blocks which wrap many of
the intricacies of the pipeline mechanics.  The library
of existing blocks also provides many of the common
operations needed for pipeline operation, both for real-time and
off-line processing.  These blocks also serve as a
practical reference for the development of new blocks. 

The low-level Python modules of Bifrost are suitable for
advanced usage or pipelines where performance is more critical.
These modules provide lightweight wrappers around the \cpp library
calls. These modules, when combined with direct access
of data through \texttt{BFarray}'s, allow for application and algorithm-specific
tuning.  Finally, at the lowest level, the \cpp
library provides an interface for developing the more performance-critical
applications.
%The \cpp library also allows for the Bifrost library into existing pipelines.
%Miles: I commented out the above because I do not understand the meaning.

In addition to increasing development productivity, Bifrost also provides
a number of monitoring points for running pipelines. 
The monitoring points are useful for not only debugging
and tuning pipelines but also for overall monitoring of
production software.  The monitoring points are implemented in
the high-level Python blocks and provide information on the
time required for acquiring data from a ring, the
amount of time spent processing the data, and the
time required to reserve a span in the output
ring.  For network related blocks, such as {\tt
bifrost.udp\_capture}, other monitoring points are exposed that profile the
packet statistics and can be used to look for
packet loss.  Bifrost also provides a collection of
UNIX-style tools for polling the monitoring points.  For
example, {\tt like\_top.py} provides a per-block analysis of the
processing load and timing in quasi-realtime.  For packet
capture monitoring the {\tt like\_bmon.py} script provides information on
any packet loss. These
are outlined in \cref{sec:tools}.
Similar to the high-level blocks,
these tools also provide working examples for the development
of application-specific monitoring software.

\subsubsection{Usability comparisons with a handwritten GPU pipeline}
\label{sec:usability_vs_gpuspec}

A Bifrost implementation of a pipeline which channelizes data
generated by the
Green Bank Ultimate Pulsar Processing Instrument (GUPPI raw data)
\footnote{See  \url{https://safe.nrao.edu/wiki/bin/view/CICADA/GUPPiUsersGuide}}
into filterbank data
(from the SIGPROC
project \footnote{See \url{http://sigproc.sourceforge.net/}})
is compared with
unmodified handwritten legacy GPU pipeline
code written as part of the
Breakthrough Listen SETI (Search for Extraterrestrial Intelligence) project
\citep[see][]{WORDEN201798,macmahon_breakthrough_2017}
software stack aboard the
GBT.

This comparison of syntax is meant to be a numerical way
to quantify ease-of-development, and makes use of the following
three metrics:

\begin{itemize}
\item Source Lines of Code (SLOC), which is the
number of lines of code which contain an executable
statement (i.e., no
comments or blanks). This measures the verbosity
required to develop the pipeline,
which is roughly correlated to the total development effort.
This measure excludes the effort required for debugging
and benchmarking code.
\item Tokens, which is the total number of words,
constants, and operators
appearing in the code. This is an alternative to SLOC
in measuring the verbosity required to develop
a pipeline, as it counts verbosity in a way
that stresses not vertical length but rather
the number of independent grammatical symbols
in a statement.
Because of how a programming habit such as
breaking up each long function call across several lines
might significantly impact the SLOC score, counting the tokens
would only account for complexity in
terms of number of words, not in terms of lines of code.
\item Cyclomatic Complexity Number (CCN) \citep[see][]{mccabe_complexity_1976},
which, in some sense,
measures the number of different paths the control flow
can move through code (i.e., entering an \textit{if} statement
or not entering an \textit{if} statement
are two different paths through code). This provides a way
to roughly quantify how difficult it will be to test
some piece of functionality. CCN can complement
the SLOC and tokens by quantifying the complexity
of code rather than the verbosity of code, which
can demonstrate how difficult it was to actually debug
and test a segment of code. CCN is measured using
the tool Lizard\footnote{\url{https://github.com/terryyin/lizard}}.
\end{itemize}

%TODO: This 7em will probably have to be adjusted in the final version.
\newcolumntype{C}{>{\centering\arraybackslash}p{4.5em}}

% Please add the following required packages to your document preamble:
% \usepackage{booktabs}
\begin{table}[tbp]
\centering
\begin{tabular}{@{}lCCCC@{}}
\cmidrule(l){2-5}
                                            & Files        & SLOC        & CCN        & Tokens        \\ \cmidrule(l){2-5} 
\multicolumn{1}{r}{\textbf{Without Bifrost:}}         & 4            & 1329        & 265        & 12638         \\
\multicolumn{1}{r}{\textbf{With Bifrost:}}            & 3            & 177         & 10         & 396           \\
\bottomrule
\multicolumn{5}{l}{SLOC - Source Lines of Code (no blanks or comments).} \\
\multicolumn{5}{l}{CCN - Cyclomatic Complexity Number, accumulated over each function.}                 \\
\multicolumn{5}{l}{Tokens - Number of words, operators, or constants.}                                \\
\end{tabular}
\caption{Comparison of a Bifrost version
of a pipeline channelizing radio data into the frequency
domain, against the original legacy pipeline. The comparison
shown is only by code syntax, with three
metrics that quantify development effort. All three metrics
are inversely correlated with development effort.
}
\label{tbl:gpuspec_comparison}
\end{table}

The scores on each metric for the pipeline
with Bifrost and for the legacy code are given
in \cref{tbl:gpuspec_comparison}. Bifrost
outclasses the legacy C/CUDA code in all categories,
having 7.5x fewer source lines of code,
26.5x smaller cyclomatic complexity number,
and 31.9x fewer number of tokens. If these metrics
can be used together to quantify development effort,
then it has been shown that Bifrost greatly eases the development
process, and makes the final pipeline far less complicated
than it had been in the hand-written version.

\section{LWA Sevilleta implementation}
\label{sec:lwa}
The Long Wavelength Array (LWA) is a proposed low-frequency
instrument sited in the southwest United States that
will provide arc-second resolution for the frequency range of
10 to 88 MHz by cross-correlating the beamformed
output of 53 stations.  The first station of
the LWA, LWA1 \cite[see][]{taylor_first_2012}
is co-located with the Karl G.
Jansky Very Large Array (VLA) in the state of New Mexico while
the second station, LWA-SV (LWA Sevilleta), is located $\sim$75 km northeast
on the Sevilleta National Wildlife Refuge.  Both stations
consist of 256 dual-polarization dipole antennas spread over a
100 by 110 m area.  LWA1 uses a
custom Field-programmable gate array-based (FPGA) digital back end while LWA-SV uses a Bifrost-based
back end running on commodity hardware.

The LWA-SV digital back end consists of 16 Reconfigurable
Open Architecture Computing Hardware version 2 (ROACH2) boards
and seven dual-GPU servers.  The ROACH2 boards each
contain two analog-to-digital converter (ADC) ADC16x256-8 digitizer cards and accept analog signals
from 16 dual-polarization antennas.  After digitization at 204.8
Ms/s, the signals are Fourier transformed into 4,096 channels
(25 kHz spectral and 40 $\mu$s time resolution), sub-banded
into three spectral tunings, and then sent to the
GPU servers.  On the servers, the data are
further processed into the three main modes of operation
for LWA-SV.

The first two modes, Transient Buffer Narrowband (TBN)
and Transient Buffer Full band (TBF), provide data
from all 256 dipoles while the final mode,
beamformed Data Receiver (DRX),
provides time domain beamformed data. These modes
are all implemented as Bifrost pipelines.
The two all-dipole
modes differ in the type of data returned and
the duty cycle.  TBN provides a 100 kHz
bandwidth time series of complex voltages with a 100\% duty
cycle which is suitable for correlation.  In contrast,
TBF provides limited duration ($<$ 1s) dumps of the
raw ``F-engine'' (converts to frequency domain)
data over a much larger bandwidth ($\sim$
20 MHz) with a lower duty cycle ($\approx$0.5\%). 
The TBF data are suitable for triggered acquisitions on
external triggers, e.g., lightning or cosmic rays.  The
beamforming mode generates complex voltage timeseries data that are
coherently summed to maximize the gain in a particular
direction.  This mode provides dual polarization data for
two spectral tunings, each of $\approx$10 MHz, that are
suitable for cross-correlation with LWA1.  

The details of the Bifrost pipelines for all three modes are provided below.  

\subsection{Packet capture implementation}
The ROACH2 boards transmit the frequency domain data to
the servers using UDP packets.  Each packet consists
of a 128-bit header that specifies basic metadata, i.e.,
the source of the packet, frequency, and time.  The data section consists
of up to 256 channels from all 16 dual-polarization
inputs to the board represented at 4+4 complex signed
integers.  Since the size of the packets scale
with bandwidth, each ROACH2 board provides a constant rate
of 25k packets/s.

The Bifrost {\tt udp\_capture} block is used to capture
these packets, unpack the payload, and reorder the packets
sent from the various ROACH2 boards.  The capture
code is optimized to use single instruction, multiple data (SIMD)
operations and the
Mellanox Messaging Accelerator (VMA) library to improve the efficiency
of the capture.  These, combined with interrupt binding
and ethernet interface tuning, allow each GPU node to
capture $\approx$13 Gb/s and 800k packets/s.  After capture
and unpacking the data are placed into a Bifrost
ring for additional processing.

\subsection{TBN pipeline}

\begin{figure}[h]
\begin{center}
\includegraphics[width=\smallfiguresize\textwidth]{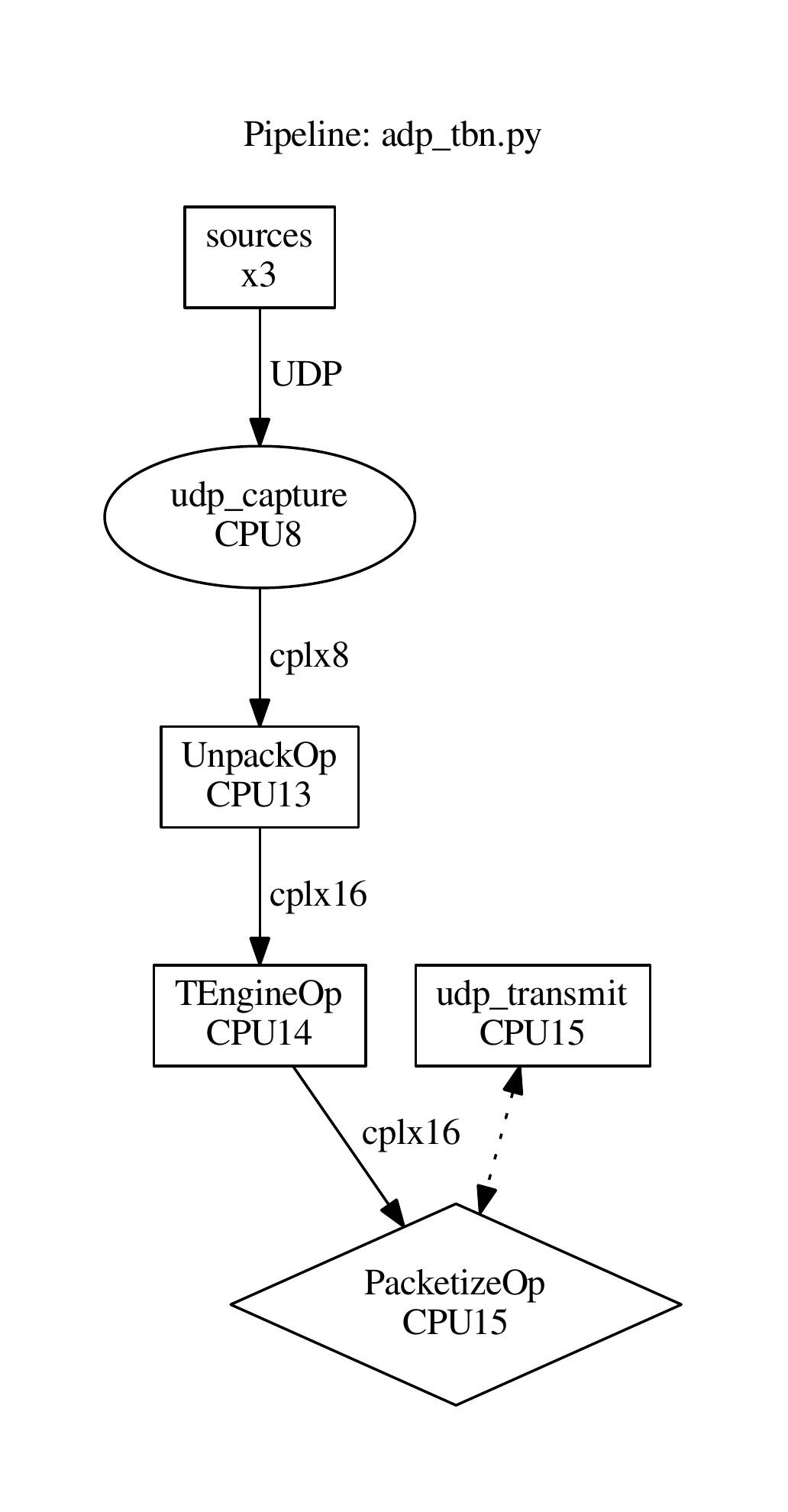}
\end{center}
\caption{TBN pipeline on the Advanced Digital Processor
at LWA-SV generated by the {\tt pipeline2dot.py}
utility included with Bifrost. 
The ovals and diamonds represent network
capture and transmit process, respectively. 
The labeled rectangles denote computing blocks.
Each block is labeled with the
operation name as well as the
CPU core binding being used. 
Finally, the labeled arrows represent the
ring buffers and the associated data
type contained within each ring.}
\label{tbn:fig}
\end{figure}

The TBN pipeline runs on six of the GPU
servers and receives 200 kHz of data from three
ROACH2 boards (48 dual-polarization antennas) at a data rate
of about 18 MB/s.  The pipeline consists of
the four stages shown in \cref{tbn:fig}:  packet
capture, unpacking, the ``T-engine'' (which converts
frequency-domain channelized data back into a time stream),
and a packetizer.  The
packet capture operation is described above.  The unpacking
stage converts the 4+4 bit complex data into a
more standard 8+8 bit complex data using {\tt bifrost.quantize}\footnote{For
historical reasons, the LWA-SV pipelines are implemented using the
low-level Python wrappers of the \cpp library functions.}. 
This 8+8 bit data is then passed to the
T-engine where it is filtered down to the 100
kHz needed for TBN.

Internally, the T-engine consists of several steps.  First,
the data is coerced to a 64-bit complex floating
point representation and then reduced in bandwidth to 100
kHz.  After the bandwidth reduction, the data are
shifted along the channel axis in preparation for the
inverse Fourier transform.  The inverse transform is carried
out on the CPU using the
{\tt scipy.fftpack.ifft}\footnote{\url{https://github.com/scipy/scipy}}
function. Once the data is in the time domain it
is combined with a complex mixer to provide sub-channel
tuning capabilities and then quantized to 8+8 bit signed
complex integers.  The quantization is performed using the
{\tt bifrost.quantize} module.  The repacked data are then
sent to the packetizer where they are formatted as
TBN data and sent to a data recorder. 
The output data rate is $\approx$18 MB/s per server.

\subsection{DRX Pipeline}

\begin{figure}[h]
\begin{center}
\includegraphics[width=\smallfiguresize\textwidth]{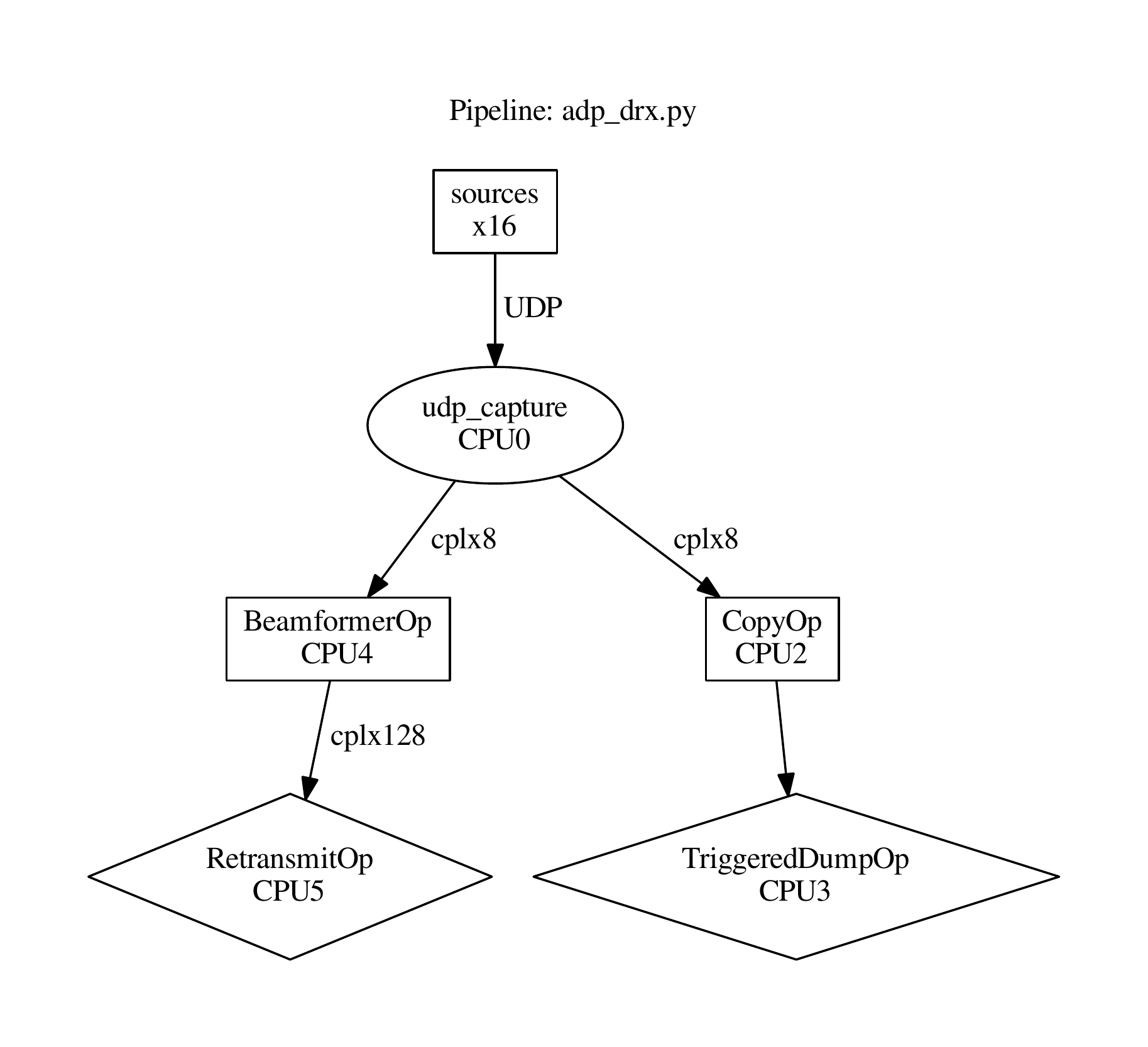}
\end{center}
\caption{DRX beamforming pipeline on the Advanced Digital Processor at LWA-SV.  The labeling is the same as in \cref{tbn:fig}.}
\label{drx:fig}
\end{figure}

Unlike TBN, the DRX pipeline captures data from all
16 ROACH2 boards.  Six of the GPU servers
runs two DRX pipelines, one for each spectral tuning
of the beamformer, that capture 10.8 MHz of bandwidth
at a data rate of $\sim$800 MB/s.  The
DRX pipeline, shown in \cref{drx:fig}, is also more
complicated than TBN since it also provides the TBF
mode. After the packet capture, the data is copied
into a secondary ring using the {\tt bifrost.memory} module
to provide the TBF data buffer.  The copy
is done in a non-guaranteed manner such in order
to preserve the state of the TBF buffer when
a dump is triggered.  Since Bifrost supports branching
pipelines the captured data is also ingested by the
GPU-based beamformer module which implements a phase-and-sum beamformer based
on the {\tt lsl.misc.beamformer} module from the LWA Software
Library \cite{dowell_long_2012}.  In order to reduce the copy bandwidth
to the GPU the beamformer directly uses the 4+4
complex signed integer data.  

\begin{figure}[h]
\begin{center}
\includegraphics[width=\smallfiguresize\textwidth]{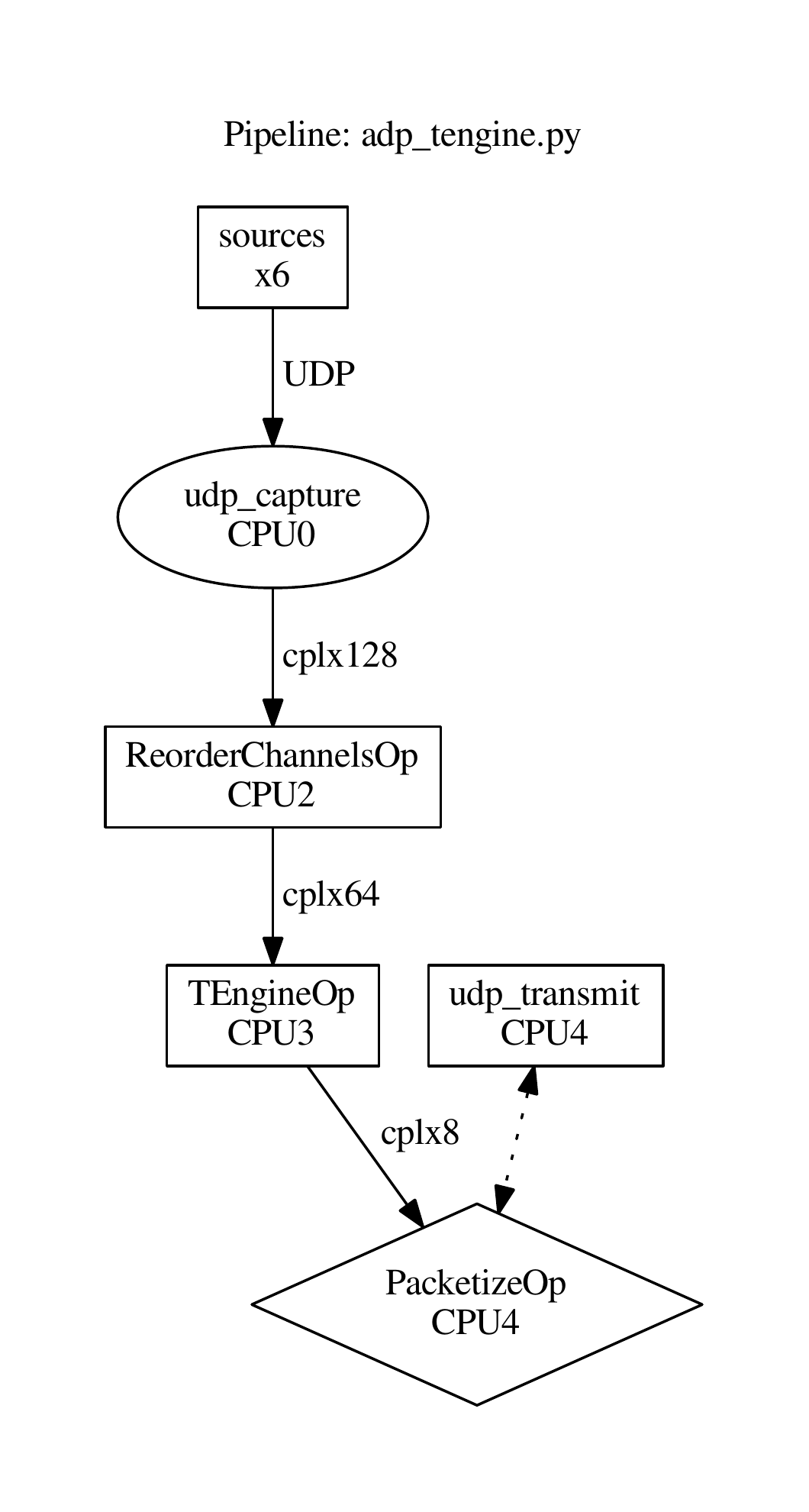}
\end{center}
\caption{DRX T-engine pipeline on the Advanced Digital Processor at LWA-SV.  The labeling is the same as in \cref{tbn:fig}.}
\label{tengine:fig}
\end{figure}

After beamforming, the data are still in the frequency
domain and need to be transformed into complex voltages.
This is done using separate T-engine pipelines, \cref{tengine:fig},
that run on the remaining seventh GPU server.  The transition to the time domain is needed for better delay compensation, i.e., less than the F-engine dump time, when cross-correlating with other LWA stations.  The time domain is also beneficial to other areas of research, particularly pulsar studies, for the higher temporal resolution.
In order to move the data from the DRX
pipelines to the T-engine pipelines the data are packetized
as UDP packets and transmitted using the {\tt bifrost.udp\_transmit}
block at a rate of about 50 MB/s per
server.  In the T-engine pipeline the data are
aggregated from all six GPU servers and transformed into
the time domain similar to TBN.  However, due
to the larger bandwidth of DRX, the inverse transform
is carried out on the GPU using the {\tt
bifrost.fft} module.  The time domain data are then
quantized to 4+4 bit complex signed integers and transmitted
to the data recorder.  The final data rate
out of the T-engine is roughly 18 MB/s per
spectral tuning.

\begin{figure}[h]
\begin{center}
\includegraphics[width=\fullfiguresize\textwidth]{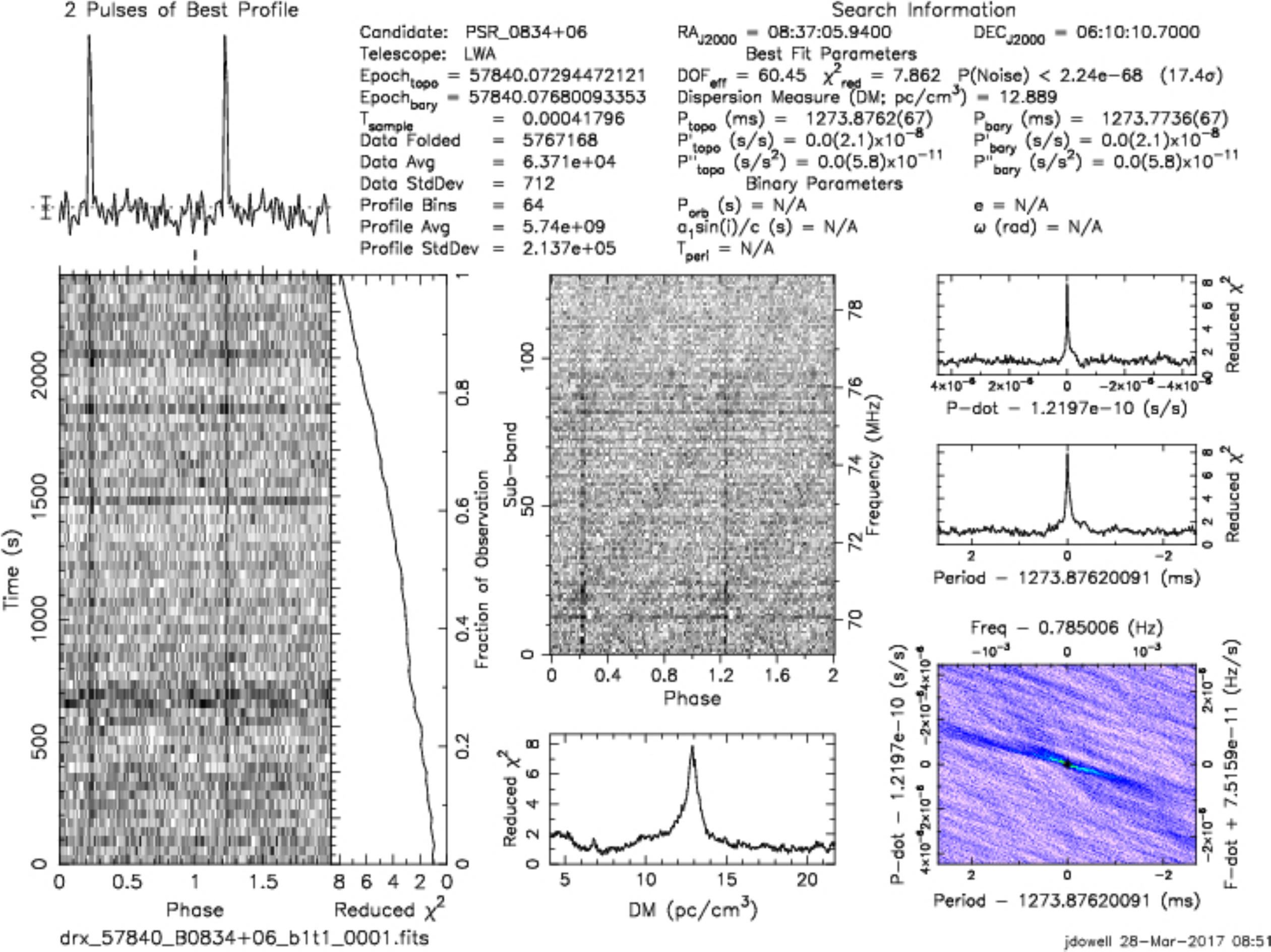}
\end{center}
\caption{Detection of the pulsar PSR B0834+06 at LWA-SV, displayed
using \textsc{PRESTO} \cite{ransom_new_2001}, the middle two panels show the pulse as a function of frequency and the best-fit dispersion measure for the observation.
This detection offers evidence that the LWA-SV pipelines, written
with Bifrost, are performing correctly.}
\label{psr:fig}
\end{figure}

To validate the Bifrost-based beamformer system, we observed the pulsar
PSR B0834+06 with LWA-SV on 2017-03-18. Due to the transient
nature of the pulsar, detection requires post-processing using a pulsar
dedispersion and folding analysis pipeline; here, we use
PRESTO\footnote{\url{https://github.com/scottransom/presto}}
\cite{ransom_new_2001}.
Successful detection of the pulsar requires system stability and
for frequency and time metadata to be correctly propagated through
the beamformer pipeline. \Cref{psr:fig} shows a detection of the
pulsar PSR B0834+06 made with one spectral tuning of the
LWA-SV beamformer. The derived value for the dispersion measure (DM)
as observed is $DM=12.889$, with period $P_\text{bary} = \SI{1273.7736(67)}{ms}$
($\chi^2_\text{red} = 7.862$ for $\text{DOF}_\text{eff} = 60.45$).
The observed values are in reasonable agreement with those listed
in the ATNF pulsar database\footnote{\url{http://www.atnf.csiro.au/research/pulsar/psrcat/}}
\citep[see][where $P_\text{bary} = \SI{1273.7682915785(7)}{ms}$ \cite{hobbs2004long},
and $DM =12.8640(4)$ \cite{stovall2015pulsar}]{2016yCat...1.102034}.
Future work involves making this process
into a real-time pulsar detection pipeline, incorporating the
existing FDMT block in Bifrost.

\section{Future work}

Bifrost is under active development and continues to accrue new features and
functionality at a rapid pace. Our near-term plans include extending its
capabilities to provide comprehensive support for correlation, beamforming,
and pulsar and transient search pipelines. In the longer term we aim to add
real-time remote monitoring and control capabilities, accessible via web-based
graphical interfaces. Such features would further reduce the cost and effort
of developing and deploying new pipelines.

\section{Conclusion}

Bifrost is an open-source
software framework for rapidly composing
processing pipelines, built to
emphasize
the high-throughput
needs of radio astronomers.
The framework comes with a built-in block and
view library
which can be used to code many simple pipelines and 
which also provides
functionality for interferometry, pulsar dedispersion and timing,
and transient search pipelines. The framework also comes
with a just-in-time compilation feature for quick
development and deployment of
CUDA kernels, and several tools for pipeline
debugging and monitoring.

Modern radio telescopes
rely on high-throughput processing to meet
real-time requirements. Such pipelines
are difficult to develop and debug.
Bifrost makes the development process easier
by abstracting away a flexible ring-buffer implementation,
and providing a high-level
API to facilitate rapid composition of pipelines
through Python. Bifrost comes with full GPU support,
and can be used to exploit GPU processing
without knowledge of CUDA. By exploiting
pipeline parallelism, Bifrost gets significantly
faster speeds for GPU computation versus regular
serial pipelines created with other packages.

Bifrost can be downloaded from 
\url{https://github.com/ledatelescope/bifrost}.
Code reference documentation and tutorials
can be found from the links given in the repository.

\ifarxiv
  \acknowledgments
\else
  \section{Acknowledgments}
\fi

Bifrost development was supported by NSF grants OCI/1060067, OIA/1125087,
AST/1139974, and AST-1106059.

Miles Cranmer would like to thank Joseph Schull
and Anna Yang for their support of his research and
for their establishment of a generous
award program at McGill University.

Construction of the LWA has
been supported by the Office
of Naval Research under Contract
N00014-07-C-0147 and by AFOSR. 
Operational support of the LWA-SV
station is provided by the Air Force Research Laboratory.

\ifarxiv

\else
  \bibliographystyle{ws-jai}
\fi
\bibliography{newbib.bib}

\begin{thebibliography}{32}
\newcommand{\enquote}[1]{``#1''}
\providecommand{\natexlab}[1]{#1}
\providecommand{\url}[1]{\texttt{#1}}
\providecommand{\urlprefix}{URL }
\expandafter\ifx\csname urlstyle\endcsname\relax
  \providecommand{\doi}[1]{doi:\discretionary{}{}{}#1}\else
  \providecommand{\doi}{doi:\discretionary{}{}{}\begingroup
  \urlstyle{rm}\Url}\fi

\bibitem[{Allen \emph{et~al.}(2013)Allen, DuPrie, Berriman, Hanisch, Mink \&
  Teuben}]{allen_astrophysics_2013}
Allen, A., DuPrie, K., Berriman, B., Hanisch, R.~J., Mink, J. \& Teuben, P.~J.
  [2013]  \enquote{Astrophysics {Source} {Code} {Library},}  (eprint:
  arXiv:1212.1916), p. 387,
  \urlprefix\url{http://adsabs.harvard.edu/abs/2013ASPC..475..387A}.

\bibitem[{Allen \emph{et~al.}(2017)Allen, DuPrie, Schmidt \&
  Ryan}]{allen_2016_2017}
Allen, A., DuPrie, K., Schmidt, J. \& Ryan, P.~W. [2017]  \enquote{2016
  {Annual} {Report} - {Astrophysics} {Source} {Code} {Library},}
  \urlprefix\url{http://ascl.net/wordpress/wp-content/uploads/2015/03/ASCL_ann_report_2017_March_final.pdf}.

\bibitem[{Bandura \emph{et~al.}(2014)Bandura, Addison, Amiri, Bond,
  Campbell-Wilson, Connor, Cliche, Davis, Deng, Denman, Dobbs, Fandino, Gibbs,
  Gilbert, Halpern, Hanna, Hincks, Hinshaw, Höfer, Klages, Landecker, Masui,
  Mena~Parra, Newburgh, Pen, Peterson, Recnik, Shaw, Sigurdson, Sitwell,
  Smecher, Smegal, Vanderlinde \& Wiebe}]{bandura_canadian_2014}
Bandura, K., Addison, G.~E., Amiri, M., Bond, J.~R., Campbell-Wilson, D.,
  Connor, L., Cliche, J.-F., Davis, G., Deng, M., Denman, N., Dobbs, M.,
  Fandino, M., Gibbs, K., Gilbert, A., Halpern, M., Hanna, D., Hincks, A.~D.,
  Hinshaw, G., Höfer, C., Klages, P., Landecker, T.~L., Masui, K., Mena~Parra,
  J., Newburgh, L.~B., Pen, U.-l., Peterson, J.~B., Recnik, A., Shaw, J.~R.,
  Sigurdson, K., Sitwell, M., Smecher, G., Smegal, R., Vanderlinde, K. \&
  Wiebe, D. [2014]  \enquote{Canadian {Hydrogen} {Intensity} {Mapping}
  {Experiment} ({CHIME}) pathfinder,}  (eprint: arXiv:1406.2288), p. 914522,
  \doi{10.1117/12.2054950},
  \urlprefix\url{http://adsabs.harvard.edu/abs/2014SPIE.9145E..22B}.

\bibitem[{Barsdell \emph{et~al.}(2010)Barsdell, Barnes \&
  Fluke}]{barsdell_analysing_2010}
Barsdell, B.~R., Barnes, D.~G. \& Fluke, C.~J. [2010]  \emph{Monthly Notices of
  the Royal Astronomical Society} \textbf{408},  1936,
  \doi{10.1111/j.1365-2966.2010.17257.x},
  \urlprefix\url{http://mnras.oxfordjournals.org/content/408/3/1936}.

\bibitem[{Berriman \& Groom(2011)}]{berriman_how_2011}
Berriman, G.~B. \& Groom, S.~L. [2011]  \emph{Commun. ACM} \textbf{54},  52,
  \doi{10.1145/2043174.2043190},
  \urlprefix\url{http://doi.acm.org/10.1145/2043174.2043190}.

\bibitem[{Cisco(2016)}]{cisco_zettabyte_2016}
Cisco [2016]  \enquote{The {Zettabyte} {Era} - {Trends} and {Analysis},}
  \urlprefix\url{http://www.cisco.com/c/en/us/solutions/collateral/service-provider/visual-networking-index-vni/vni-hyperconnectivity-wp.html}.

\bibitem[{Dowell \emph{et~al.}(2012)Dowell, Wood, Stovall, Ray, Clarke \&
  Taylor}]{dowell_long_2012}
Dowell, J., Wood, D., Stovall, K., Ray, P.~S., Clarke, T. \& Taylor, G. [2012]
  \emph{Journal of Astronomical Instrumentation} \textbf{1},  1250006,
  \doi{10.1142/S2251171712500067},
  \urlprefix\url{http://adsabs.harvard.edu/abs/2012JAI.....150006D}.

\bibitem[{Ellson \emph{et~al.}(2001)Ellson, Gansner, Koutsofios, North \&
  Woodhull}]{ellson_graphviz_2001}
Ellson, J., Gansner, E., Koutsofios, L., North, S.~C. \& Woodhull, G. [2001]
  \enquote{Graphviz— {Open} {Source} {Graph} {Drawing} {Tools},}  \emph{Graph
  {Drawing}} (Springer, Berlin, Heidelberg), p. 483,
  \urlprefix\url{http://link.springer.com/chapter/10.1007/3-540-45848-4_57},
  dOI: 10.1007/3-540-45848-4\_57.

\bibitem[{Givon \emph{et~al.}(2015)Givon, Unterthiner, Erichson, Chiang,
  Larson, Pfister, Dieleman, Lee, van~der Walt, Moldovan, Bastien, jz1536, Shi,
  Schlüter, Thomas, Capdevila \& Rubinsteyn}]{lev_givon_2015_40565}
Givon, L., Unterthiner, T., Erichson, N.~B., Chiang, D.~W., Larson, E.,
  Pfister, L.~A., Dieleman, S., Lee, G.~R., van~der Walt, S., Moldovan, T.~M.,
  Bastien, F., jz1536, Shi, X., Schlüter, J., Thomas, B., Capdevila, C. \&
  Rubinsteyn, A. [2015]  \enquote{scikit-cuda 0.5.1,}
  \doi{10.5281/zenodo.40565},
  \urlprefix\url{https://doi.org/10.5281/zenodo.40565}.

\bibitem[{Hickish \emph{et~al.}(2016)Hickish, Abdurashidova, Ali, Buch,
  Chaudhari, Chen, Dexter, Domagalski, Ford, Foster, George, Greenberg,
  Greenhill, Isaacson, Jiang, Jones, Kapp, Kriel, Lacasse, Lutomirski,
  MacMahon, Manley, Martens, McCullough, Muley, New, Parsons, Price, Primiani,
  Ray, Siemion, Van~Tonder, Vertatschitsch, Wagner, Weintroub \&
  Werthimer}]{hickish2016}
Hickish, J., Abdurashidova, Z., Ali, Z., Buch, K.~D., Chaudhari, S.~C., Chen,
  H., Dexter, M., Domagalski, R.~S., Ford, J., Foster, G., George, D.,
  Greenberg, J., Greenhill, L., Isaacson, A., Jiang, H., Jones, G., Kapp, F.,
  Kriel, H., Lacasse, R., Lutomirski, A., MacMahon, D., Manley, J., Martens,
  A., McCullough, R., Muley, M.~V., New, W., Parsons, A., Price, D.~C.,
  Primiani, R.~A., Ray, J., Siemion, A., Van~Tonder, V., Vertatschitsch, L.,
  Wagner, M., Weintroub, J. \& Werthimer, D. [2016]  \emph{Journal of
  Astronomical Instrumentation} \textbf{05},  1641001,
  \doi{10.1142/S2251171716410014},
  \urlprefix\url{http://www.worldscientific.com/doi/abs/10.1142/S2251171716410014}.

\bibitem[{Hobbs \emph{et~al.}(2004)Hobbs, Lyne, Kramer, Martin \&
  Jordan}]{hobbs2004long}
Hobbs, G., Lyne, A., Kramer, M., Martin, C. \& Jordan, C. [2004]  \emph{Monthly
  Notices of the Royal Astronomical Society} \textbf{353},  1311.

\bibitem[{Jones \emph{et~al.}(2012)Jones, Wagstaff, Thompson, D'Addario,
  Navarro, Mattmann, Majid, Lazio, Preston \& Rebbapragada}]{jones2012}
Jones, D.~L., Wagstaff, K., Thompson, D.~R., D'Addario, L., Navarro, R.,
  Mattmann, C., Majid, W., Lazio, J., Preston, R. \& Rebbapragada, U. [2012]
  \enquote{Big data challenges for large radio arrays,}  \emph{2012 IEEE
  Aerospace Conference}, p. 1, \doi{10.1109/AERO.2012.6187090}.

\bibitem[{Klöckner \emph{et~al.}(2012)Klöckner, Pinto, Lee, Catanzaro, Ivanov
  \& Fasih}]{klockner_pycuda_2012}
Klöckner, A., Pinto, N., Lee, Y., Catanzaro, B., Ivanov, P. \& Fasih, A.
  [2012]  \emph{Parallel Comput.} \textbf{38},  157,
  \doi{10.1016/j.parco.2011.09.001},
  \urlprefix\url{http://dx.doi.org/10.1016/j.parco.2011.09.001}.

\bibitem[{Kocz \emph{et~al.}(2015)Kocz, Greenhill, Barsdell, Price, Bernardi,
  Bourke, Clark, Craig, Dexter, Dowell, Eftekhari, Ellingson, Hallinan,
  Hartman, Jameson, MacMahon, Taylor, Schinzel \&
  Werthimer}]{kocz_digital_2015}
Kocz, J., Greenhill, L.~J., Barsdell, B.~R., Price, D., Bernardi, G., Bourke,
  S., Clark, M.~A., Craig, J., Dexter, M., Dowell, J., Eftekhari, T.,
  Ellingson, S., Hallinan, G., Hartman, J., Jameson, A., MacMahon, D., Taylor,
  G., Schinzel, F. \& Werthimer, D. [2015]  \emph{Journal of Astronomical
  Instrumentation} \textbf{4},  1550003, \doi{10.1142/S2251171715500038},
  \urlprefix\url{http://adsabs.harvard.edu/abs/2015JAI.....450003K}.

\bibitem[{Lupton \emph{et~al.}(2001)Lupton, Gunn, Ivezić, Knapp \&
  Kent}]{lupton_sdss_2001}
Lupton, R., Gunn, J.~E., Ivezić, Z., Knapp, G.~R. \& Kent, S. [2001]
  \enquote{The {SDSS} {Imaging} {Pipelines},}  (eprint:
  arXiv:astro-ph/0101420), p. 269,
  \urlprefix\url{http://adsabs.harvard.edu/abs/2001ASPC..238..269L}.

\bibitem[{MacMahon \emph{et~al.}(2017)MacMahon, Price, Lebofsky, Siemion,
  Croft, DeBoer, Enriquez, Gajjar, Hellbourg, Isaacson, Werthimer,
  Abdurashidova, Bloss, Creager, Ford, Lynch, Maddalena, McCullough, Ray,
  Whitehead \& Woody}]{macmahon_breakthrough_2017}
MacMahon, D. H.~E., Price, D.~C., Lebofsky, M., Siemion, A. P.~V., Croft, S.,
  DeBoer, D., Enriquez, J.~E., Gajjar, V., Hellbourg, G., Isaacson, H.,
  Werthimer, D., Abdurashidova, Z., Bloss, M., Creager, R., Ford, J., Lynch,
  R.~S., Maddalena, R.~J., McCullough, R., Ray, J., Whitehead, M. \& Woody, D.
  [2017]  \emph{arXiv:1707.06024 [astro-ph]}
  \urlprefix\url{http://arxiv.org/abs/1707.06024}, arXiv: 1707.06024.

\bibitem[{{Manchester} \emph{et~al.}(2016){Manchester}, {Hobbs}, {Teoh} \&
  {Hobbs}}]{2016yCat...1.102034}
{Manchester}, R.~N., {Hobbs}, G.~B., {Teoh}, A. \& {Hobbs}, M. [2016]
  \emph{VizieR Online Data Catalog} .

\bibitem[{McCabe(1976)}]{mccabe_complexity_1976}
McCabe, T.~J. [1976]  \emph{IEEE Transactions on Software Engineering}
  \textbf{SE-2},  308, \doi{10.1109/TSE.1976.233837}.

\bibitem[{Messick \emph{et~al.}(2017)Messick, Blackburn, Brady, Brockill,
  Cannon, Cariou, Caudill, Chamberlin, Creighton, Everett, Hanna, Keppel, Lang,
  Li, Meacher, Nielsen, Pankow, Privitera, Qi, Sachdev, Sadeghian, Singer,
  Thomas, Wade, Wade, Weinstein \& Wiesner}]{messick_analysis_2017}
Messick, C., Blackburn, K., Brady, P., Brockill, P., Cannon, K., Cariou, R.,
  Caudill, S., Chamberlin, S.~J., Creighton, J. D.~E., Everett, R., Hanna, C.,
  Keppel, D., Lang, R.~N., Li, T. G.~F., Meacher, D., Nielsen, A., Pankow, C.,
  Privitera, S., Qi, H., Sachdev, S., Sadeghian, L., Singer, L., Thomas, E.~G.,
  Wade, L., Wade, M., Weinstein, A. \& Wiesner, K. [2017]  \emph{Physical
  Review D} \textbf{95},  042001, \doi{10.1103/PhysRevD.95.042001},
  \urlprefix\url{http://adsabs.harvard.edu/abs/2017PhRvD..95d2001M}.

\bibitem[{Mort \emph{et~al.}(2015)Mort, Dulwich, Williams \&
  Salvini}]{mort_pelican:_2015}
Mort, B., Dulwich, F., Williams, C. \& Salvini, S. [2015]  \emph{Astrophysics
  Source Code Library} ,
  ascl:1507.003\urlprefix\url{http://adsabs.harvard.edu/abs/2015ascl.soft07003M}.

\bibitem[{Price(2016)}]{price_spectrometers_2016}
Price, D.~C. [2016]  \emph{ArXiv e-prints} \textbf{1607},  arXiv:1607.03579,
  \urlprefix\url{http://adsabs.harvard.edu/abs/2016arXiv160703579P}.

\bibitem[{Price \emph{et~al.}(2016)Price, Staveley-Smith, Bailes, Carretti,
  Jameson, Jones, van Straten \& Schediwy}]{price_hipsr:_2016}
Price, D.~C., Staveley-Smith, L., Bailes, M., Carretti, E., Jameson, A., Jones,
  M.~E., van Straten, W. \& Schediwy, S.~W. [2016]  \emph{Journal of
  Astronomical Instrumentation} \textbf{5},  1641007,
  \doi{10.1142/S2251171716410075},
  \urlprefix\url{http://adsabs.harvard.edu/abs/2016JAI.....541007P}.

\bibitem[{Quinn \emph{et~al.}(2015)Quinn, Axelrod, Bird, Dodson, Szalay \&
  Wicenec}]{quinn_delivering_2015}
Quinn, P., Axelrod, T., Bird, I., Dodson, R., Szalay, A. \& Wicenec, A. [2015]
  \emph{Advancing Astrophysics with the Square Kilometre Array (AASKA14)} ,
  147\urlprefix\url{http://adsabs.harvard.edu/abs/2015aska.confE.147Q}.

\bibitem[{Ransom(2001)}]{ransom_new_2001}
Ransom, S.~M. [2001]  \enquote{New {Search} {Techniques} for {Binary}
  {Pulsars},} p. 119.03,
  \urlprefix\url{http://adsabs.harvard.edu/abs/2001AAS...19911903R}.

\bibitem[{Ransom \emph{et~al.}(2009)Ransom, Demorest, Ford, McCullough, Ray,
  DuPlain \& Brandt}]{ransom_guppi:_2009}
Ransom, S.~M., Demorest, P., Ford, J., McCullough, R., Ray, J., DuPlain, R. \&
  Brandt, P. [2009]  \enquote{{GUPPI}: {Green} {Bank} {Ultimate} {Pulsar}
  {Processing} {Instrument},} p. 605.08,
  \urlprefix\url{http://adsabs.harvard.edu/abs/2009AAS...21460508R}.

\bibitem[{Recnik \emph{et~al.}(2015)Recnik, Bandura, Denman, Hincks, Hinshaw,
  Klages, Pen \& Vanderlinde}]{recnik_efficient_2015}
Recnik, A., Bandura, K., Denman, N., Hincks, A.~D., Hinshaw, G., Klages, P.,
  Pen, U.-L. \& Vanderlinde, K. [2015]  \emph{ArXiv e-prints} \textbf{1503},
  arXiv:1503.06189,
  \urlprefix\url{http://adsabs.harvard.edu/abs/2015arXiv150306189R}.

\bibitem[{Serylak \emph{et~al.}(2013)Serylak, Karastergiou, Williams, Armour,
  Giles \& {LOFAR Pulsar Working Group}}]{serylak_observations_2013}
Serylak, M., Karastergiou, A., Williams, C., Armour, W., Giles, M. \& {LOFAR
  Pulsar Working Group} [2013]  \enquote{Observations of transients and pulsars
  with {LOFAR} international stations and the {ARTEMIS} backend,}  (eprint:
  arXiv:1210.4318), p. 492, \doi{10.1017/S1743921312024623},
  \urlprefix\url{http://adsabs.harvard.edu/abs/2013IAUS..291..492S}.

\bibitem[{Stovall \emph{et~al.}(2015)Stovall, Ray, Blythe, Dowell, Eftekhari,
  Garcia, Lazio, McCrackan, Schinzel \& Taylor}]{stovall2015pulsar}
Stovall, K., Ray, P., Blythe, J., Dowell, J., Eftekhari, T., Garcia, A., Lazio,
  T., McCrackan, M., Schinzel, F. \& Taylor, G. [2015]  \emph{The Astrophysical
  Journal} \textbf{808},  156.

\bibitem[{Taylor \emph{et~al.}(2012)Taylor, Ellingson, Kassim, Craig, Dowell,
  Wolfe, Hartman, Bernardi, Clarke, Cohen, Dalal, Erickson, Hicks, Greenhill,
  Jacoby, Lane, Lazio, Mitchell, Navarro, Ord, Pihlström, Polisensky, Ray,
  Rickard, Schinzel, Schmitt, Sigman, Soriano, Stewart, Stovall, Tremblay,
  Wang, Weiler, White \& Wood}]{taylor_first_2012}
Taylor, G.~B., Ellingson, S.~W., Kassim, N.~E., Craig, J., Dowell, J., Wolfe,
  C.~N., Hartman, J., Bernardi, G., Clarke, T., Cohen, A., Dalal, N.~P.,
  Erickson, W.~C., Hicks, B., Greenhill, L.~J., Jacoby, B., Lane, W., Lazio,
  J., Mitchell, D., Navarro, R., Ord, S.~M., Pihlström, Y., Polisensky, E.,
  Ray, P.~S., Rickard, L.~J., Schinzel, F.~K., Schmitt, H., Sigman, E.,
  Soriano, M., Stewart, K.~P., Stovall, K., Tremblay, S., Wang, D., Weiler,
  K.~W., White, S. \& Wood, D.~L. [2012]  \emph{Journal of Astronomical
  Instrumentation} \textbf{1},  1250004, \doi{10.1142/S2251171712500043},
  \urlprefix\url{http://adsabs.harvard.edu/abs/2012JAI.....150004T}.

\bibitem[{Walt \emph{et~al.}(2011)Walt, Colbert \& Varoquaux}]{walt_numpy_2011}
Walt, S. v.~d., Colbert, S.~C. \& Varoquaux, G. [2011]  \emph{Computing in
  Science \& Engineering} \textbf{13},  22, \doi{10.1109/MCSE.2011.37},
  \urlprefix\url{http://aip.scitation.org/doi/abs/10.1109/MCSE.2011.37}.

\bibitem[{Worden \emph{et~al.}(2017)Worden, Drew, Siemion, Werthimer, DeBoer,
  Croft, MacMahon, Lebofsky, Isaacson, Hickish, Price, Gajjar \&
  Wright}]{WORDEN201798}
Worden, S.~P., Drew, J., Siemion, A., Werthimer, D., DeBoer, D., Croft, S.,
  MacMahon, D., Lebofsky, M., Isaacson, H., Hickish, J., Price, D., Gajjar, V.
  \& Wright, J.~T. [2017]  \emph{Acta Astronautica} \textbf{139},  98 ,
  \doi{http://dx.doi.org/10.1016/j.actaastro.2017.06.008},
  \urlprefix\url{http://www.sciencedirect.com/science/article/pii/S0094576517303144}.

\bibitem[{Zackay \& Ofek(2014)}]{zackay_accurate_2014}
Zackay, B. \& Ofek, E.~O. [2014]  \emph{ArXiv e-prints} \textbf{1411},
  arXiv:1411.5373,
  \urlprefix\url{http://adsabs.harvard.edu/abs/2014arXiv1411.5373Z}.

\end{thebibliography}

%\appendix

\begin{comment}
\section{Appendix}

\subsection{Bifrost Module Listing}
\end{comment}

\end{document}